\documentclass[12pt]{article}
\usepackage{epsfig}

\newcommand\pubnumber{FERMILAB-Conf-00/333-E}
\newcommand\pubdate{\today}
\newcommand\hepnumber{hep-ex/0012061}

\def\fnal{Fermi National Accelerator Laboratory\\
Batavia, IL 60510 USA}
\def\email{\footnote{womersley@fnal.gov}}

\def\Title#1{\begin{center} {\Large\bf #1 } \end{center}}
\def\Author#1{\begin{center}{ \sc #1} \end{center}}
\def\Address#1{\begin{center}{ \it #1} \end{center}}

\newcommand\pubblock{\rightline{\begin{tabular}{l} \pubnumber\\
         \pubdate\\ \hepnumber \end{tabular}}}
\newenvironment{Abstract}{\begin{quotation}  }{\end{quotation}}
\newenvironment{Presented}{\begin{quotation} \begin{center} 
             Presented at the\end{center}
      \begin{center}\begin{large}}{\end{large}\end{center} \end{quotation}}
\def\Acknowledgments{\bigskip  \bigskip \begin{center}
          \large\bf Acknowledgments\end{center}}

\makeatletter
\def\section{\@startsection{section}{0}{\z@}{5.5ex plus .5ex minus
 1.5ex}{2.3ex plus .2ex}{\large\bf}}
\def\subsection{\@startsection{subsection}{1}{\z@}{3.5ex plus .5ex minus
 1.5ex}{1.3ex plus .2ex}{\normalsize\bf}}
\def\subsubsection{\@startsection{subsubsection}{2}{\z@}{-3.5ex plus
-1ex minus  -.2ex}{2.3ex plus .2ex}{\normalsize\sl}}

\renewcommand{\@makecaption}[2]{%
   \vskip 10pt
   \setbox\@tempboxa\hbox{\small #1: #2}
   \ifdim \wd\@tempboxa >\hsize     
       \small #1: #2\par          
     \else                        
       \hbox to\hsize{\hfil\box\@tempboxa\hfil}
   \fi}

 \def\citenum#1{{\def\@cite##1##2{##1}\cite{#1}}}
 
\newcount\@tempcntc
\def\@citex[#1]#2{\if@filesw\immediate\write\@auxout{\string\citation{#2}}\fi
  \@tempcnta\z@\@tempcntb\m@ne\def\@citea{}\@cite{\@for\@citeb:=#2\do
    {\@ifundefined
       {b@\@citeb}{\@citeo\@tempcntb\m@ne\@citea\def\@citea{,}{\bf ?}\@warning
       {Citation `\@citeb' on page \thepage \space undefined}}%
    {\setbox\z@\hbox{\global\@tempcntc0\csname b@\@citeb\endcsname\relax}%
     \ifnum\@tempcntc=\z@ \@citeo\@tempcntb\m@ne
       \@citea\def\@citea{,}\hbox{\csname b@\@citeb\endcsname}%
     \else
      \advance\@tempcntb\@ne
      \ifnum\@tempcntb=\@tempcntc
      \else\advance\@tempcntb\m@ne\@citeo
      \@tempcnta\@tempcntc\@tempcntb\@tempcntc\fi\fi}}\@citeo}{#1}}
\def\@citeo{\ifnum\@tempcnta>\@tempcntb\else\@citea\def\@citea{,}%
  \ifnum\@tempcnta=\@tempcntb\the\@tempcnta\else
  {\advance\@tempcnta\@ne\ifnum\@tempcnta=\@tempcntb \else\def\@citea{--}\fi
    \advance\@tempcnta\m@ne\the\@tempcnta\@citea\the\@tempcntb}\fi\fi}
\makeatother

\def\beq{\begin{equation}}
\def\eeq#1{\label{#1}\end{equation}}
\def\eeqn{\end{equation}}

\newenvironment{Eqnarray}%
   {\arraycolsep 0.14em\begin{eqnarray}}{\end{eqnarray}}
\def\beqa{\begin{Eqnarray}}
\def\eeqa#1{\label{#1}\end{Eqnarray}}
\def\eeqan{\end{Eqnarray}}

\let\bar=\overbar

\def\O{{\cal O}}

\def\Dslash{\not{\hbox{\kern-4pt $D$}}}
\def\dslash{\not{\hbox{\kern-2pt $\del$}}}

\def\msb{{\bar{\ssstyle M \kern -1pt S}}}

\def\lsim{\mathrel{\raise.3ex\hbox{$<$\kern-.75em\lower1ex\hbox{$\sim$}}}}
\def\gsim{\mathrel{\raise.3ex\hbox{$>$\kern-.75em\lower1ex\hbox{$\sim$}}}}

\begin{document}
\begin{titlepage}
\pubblock

\vfill
\def\thefootnote{\fnsymbol{footnote}}
\Title{QCD at the Tevatron: \\[5pt] Status and Prospects}
\vfill
\Author{John Womersley\email}
\Address{\fnal}
\vfill
\begin{Abstract}
I shall review the present status of Tevatron QCD studies, focusing
on the production of jets, vector bosons, photons and heavy quarks.
In general there is good agreement between the results of current 
calculational tools and the experimental data. 
The major areas of discrepancy
arise when the input parton distributions become uncertain (for
example, jets at high $E_T$) or when the momentum scales become
relatively small (for example, $b$ production at low $p_T$).  
We can look forward to continued improvement in both calculations
and measurements over the next decade.  However, fully exploiting 
the power of the data will require considerable work,
both from the experimentalists who must
understand and publish all the systematic errors and their
correlations, and from the phenomenologists who must
understand the level of uncertainty in their calculations and 
in the parton distributions.
\end{Abstract}
\vfill
\begin{Presented}
5th International Symposium on Radiative Corrections \\ 
(RADCOR--2000) \\[4pt]
Carmel CA, USA, 11--15 September, 2000
\end{Presented}
\vfill
\end{titlepage}
\def\thefootnote{\arabic{footnote}}
\setcounter{footnote}{0}


\section{Introduction}

\begin{figure}[tb]
\begin{center}
\includegraphics*[height=6cm]{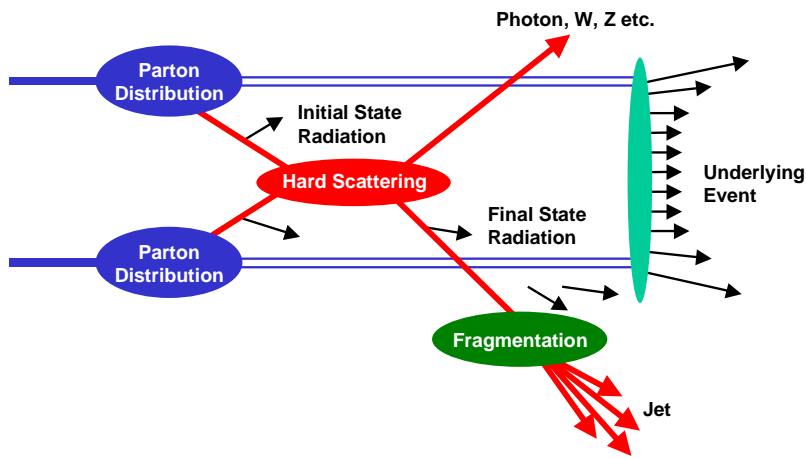}
\caption{Schematic view of a hadron-hadron collision.
\label{fig:collision}}
\end{center}
\end{figure}

\begin{figure}[tb]
\begin{center}
\includegraphics*[bb=60 80 600 700, height=7cm, angle=270]{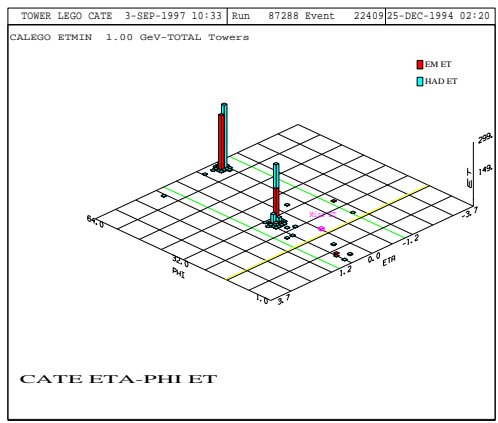}
\caption{Lego plot of a high-$E_T$ dijet event in D\O. 
Towers with $E_T<1$~GeV are suppressed.
\label{fig:lego}}
\end{center}
\end{figure}

It is over four years since data taking was completed in Run~1 
at the Tevatron, so there are rather few new results to report.
Instead, this presentation will be more of a review of the current
state of knowledge, highlighting unresolved issues and prospects 
for Run~2.  I shall briefly cover the production of jets, vector 
bosons, photons and heavy flavor.  The opinions expressed are
my own, and do not necessarily reflect the ``party line'' of the
experiments.

The Tevatron collider recorded about 100~pb$^{-1}$ of data during
1992--95 (Run~1), with two large detctors, CDF and D\O.  The 
results I shall show all come from this dataset, which was taken at 
$\sqrt{s}=1.8$~TeV (with a small amount of running at 630~GeV).  
The detectors are now nearing the completion of major upgrades, 
and data taking will resume in March 2001 (Run~2). The goal is 
to accumulate 2~fb$^{-1}$ by 2003 and 15~fb$^{-1}$ by 2007. 
In Run~2, the machine will operate at $\sqrt{s}=1.96$~TeV.

In hadron-hadron collisions, 
the simple picture of perturbative
hard scattering between point-like particles, as described 
for $e^+e^-$ collisions by earlier speakers at this meeting,  
becomes complicated by the additional effects shown in 
Fig.~\ref{fig:collision}:
\begin{itemize}
\item parton distributions --- a hadron collider is really a 
broad-band quark and gluon collider;
\item fragmentation of final state quarks and gluons;
\item both the initial and final states can be colored and
can radiate gluons, which may interfere;
\item the presence of an underlying event from proton remnants.
\end {itemize}
Despite these potential complications, at sufficiently high energies 
the events 
appear quite simple: clear two-jet structure becomes obvious, as
seen in Fig.~\ref{fig:lego}, for example. 
Let us start by reviewing the status of jet production.

\section {Jet Production}

\subsection{Inclusive Jet Cross sections at $\sqrt{s}=1.8$~TeV}

\begin{figure}[p]
\begin{center}
\begin{tabular}{cc}
\includegraphics*[bb=30 140 525 655,height=6cm]{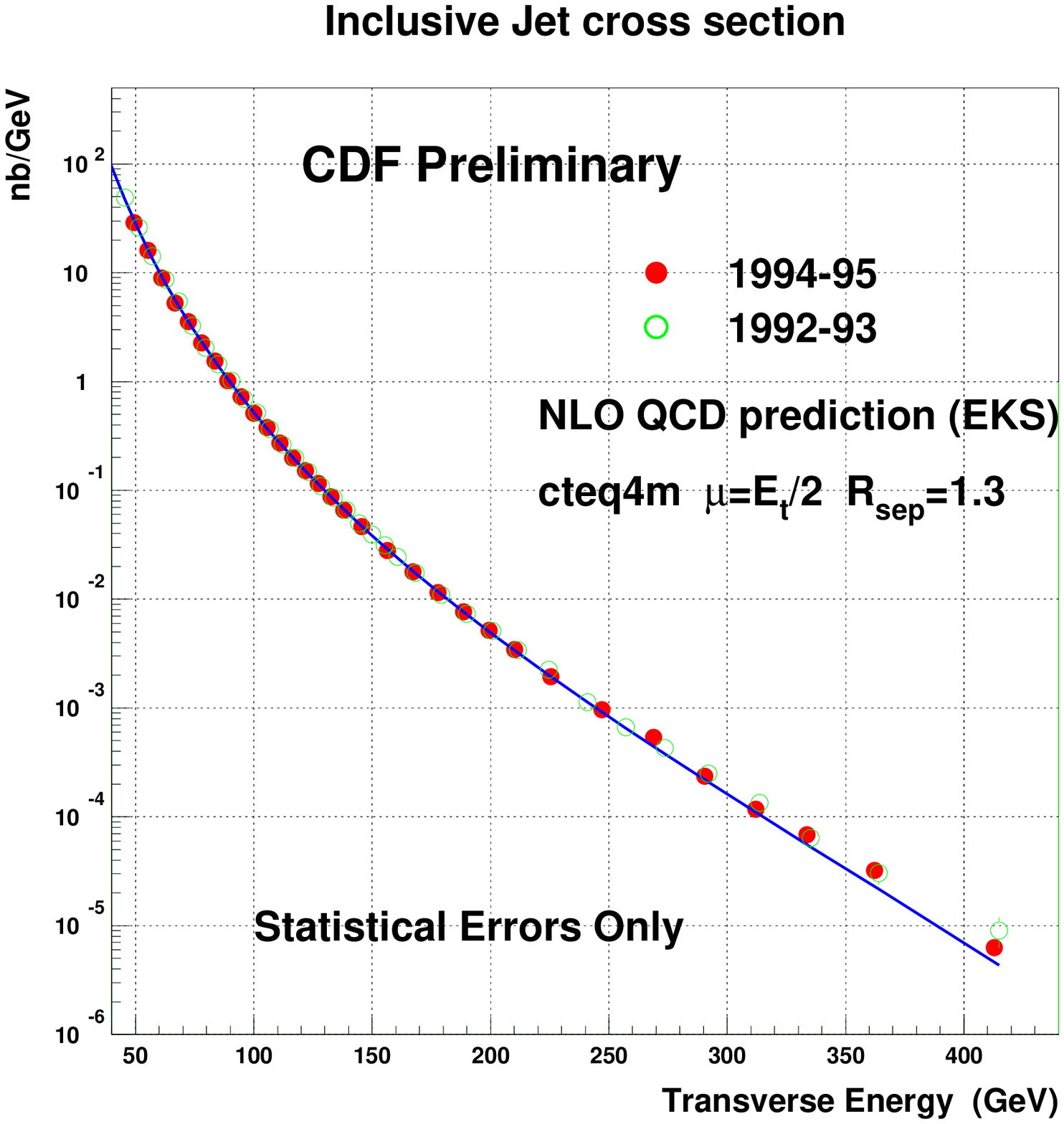}&
\includegraphics*[height=6cm]{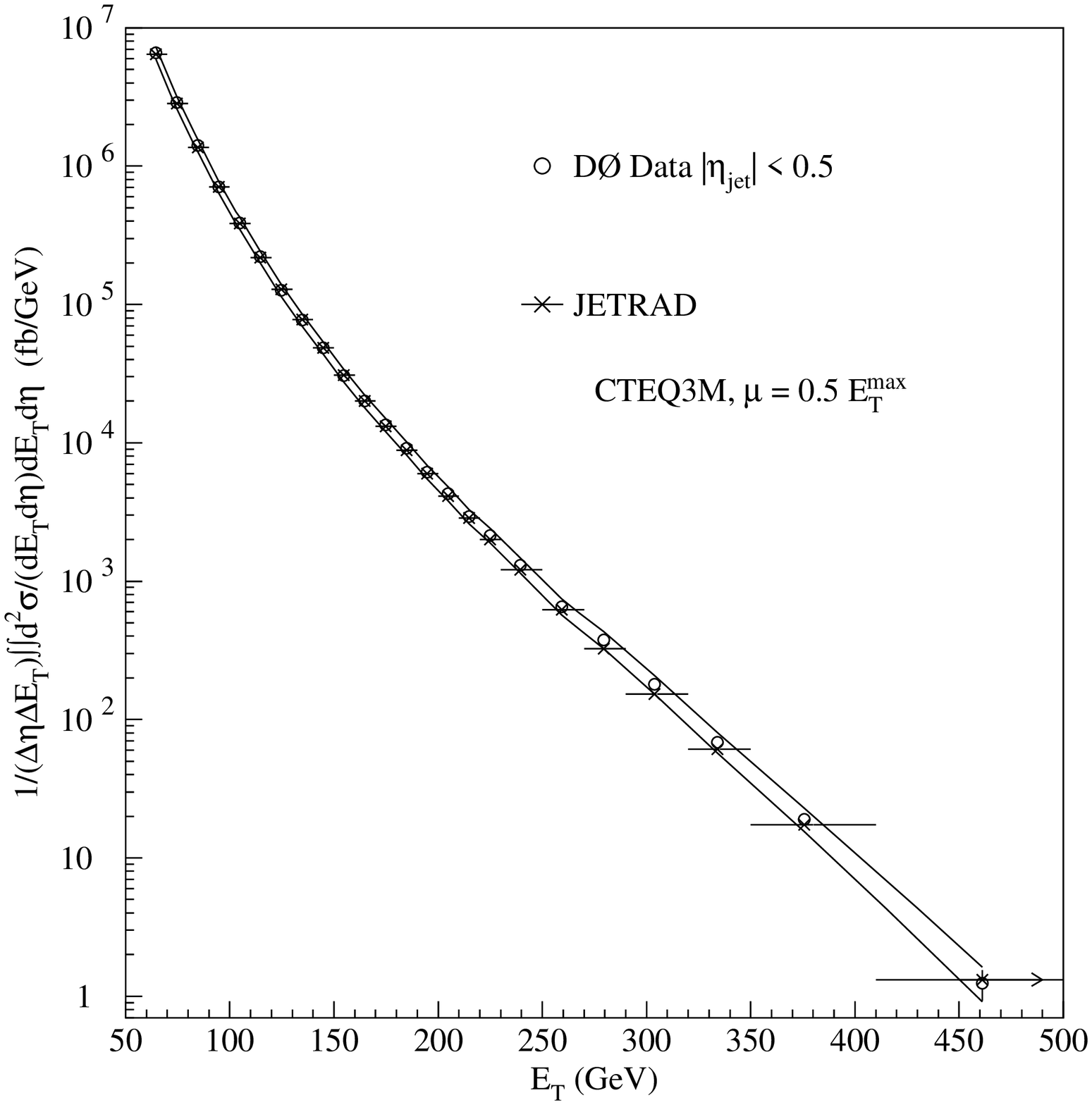}\\
\end{tabular}
\caption{Inclusive jet cross sections measured at the
Tevatron by CDF\protect\cite{cdfjets} 
(left, for $0.1<|\eta|<0.7$) and D\O\protect\cite{d0jets} 
(right, for $|\eta|< 0.5$), compared to the NLO QCD prediction.}
\label{fig:tevlogjets}
\vspace{2cm}
\begin{tabular}{cc}
\includegraphics*[bb=30 140 525 655,height=6cm]{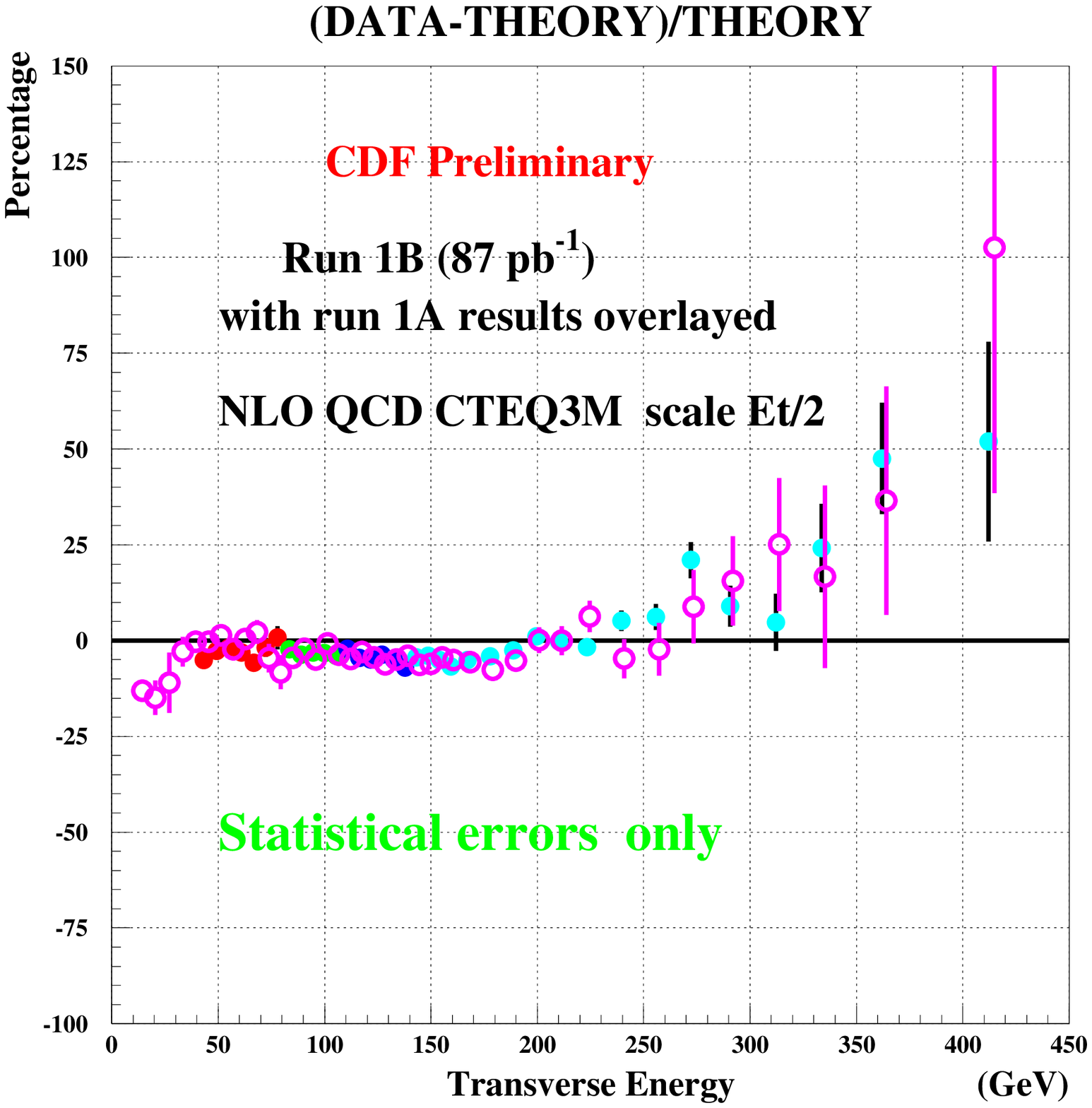}&
\includegraphics*[height=6cm]{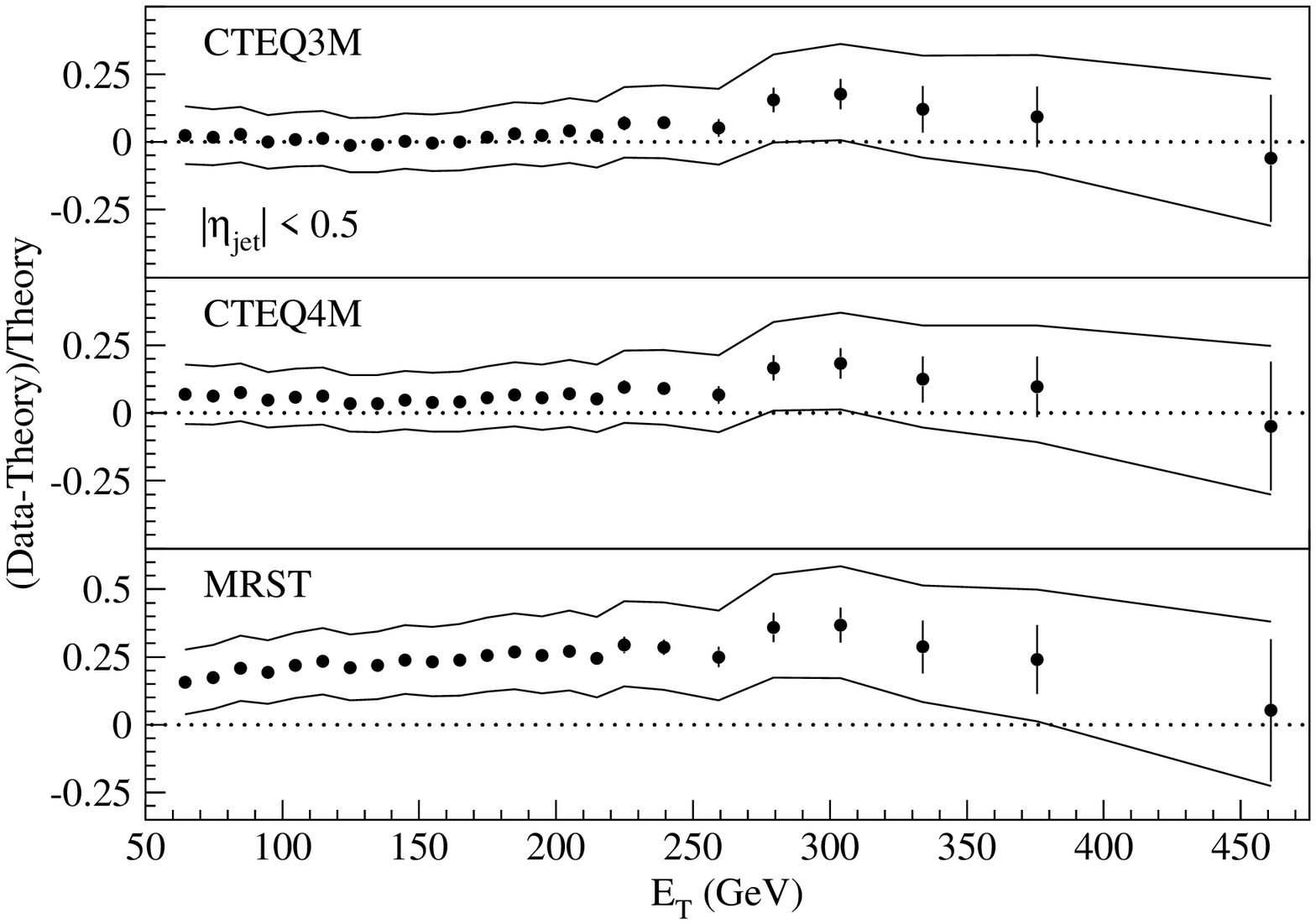}\\
\end{tabular}
\caption{Inclusive jet cross sections measured at the
Tevatron by CDF\protect\cite{cdfjets} 
(left, for $0.1<|\eta|<0.7$) and D\O\protect\cite{d0jets} 
(right, for $|\eta|< 0.5$), all normalized to the NLO QCD prediction.}
\label{fig:tevjets}
\end{center}
\end{figure}

CDF\cite{cdfjets} and D\O\cite{d0jets} 
have both measured the cross section for $R=0.7$ cone
jets in the central rapidity region.  The cross section falls by seven 
orders of magnitude between $E_T=50$ and 450~GeV and both 
experiments' data are in pretty
good agreement with NLO QCD over the whole range, as
seen in Fig.~\ref{fig:tevlogjets}.  Looked at on
a linear scale and normalized to the prediction, however, we
have the situation shown in Fig.~\ref{fig:tevjets}
(note that the CDF figure does not include systematic errors).
The impression one gets is that there is a marked excess above
QCD in the CDF data, which is not observed at D\O.
So much has been said about this discrepancy
that it is difficult to know what
can usefully be added\footnote{see Fig.~1 in \cite{jwlp99}.} but
I shall attempt to describe where we now stand.  

\begin{figure}[tb]
\begin{center}
\includegraphics*[height=8cm]{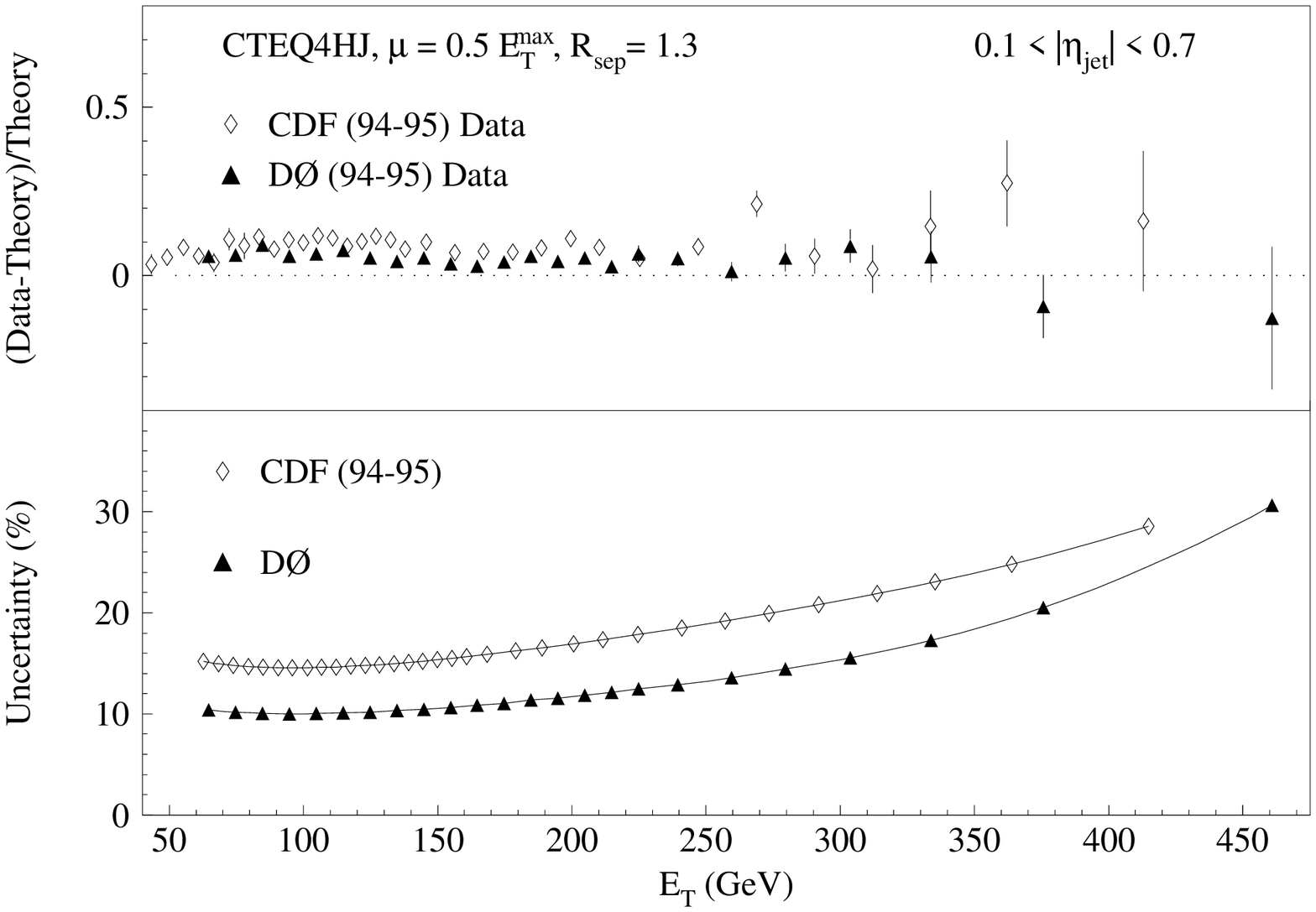}
\vspace{-2cm}
\caption{Inclusive jet cross sections for $0.1<|\eta|<0.7$
from CDF and D\O, compared with CTEQ4HJ distribution; and
size of the systematic errors on the two measurements.
Taken from \protect\cite{annrev}.} 
\label{fig:annrev}
\end{center}
\end{figure}

\begin{figure}[tb]
\begin{center}
\begin{tabular}{cc}
\includegraphics*[height=5.7cm]{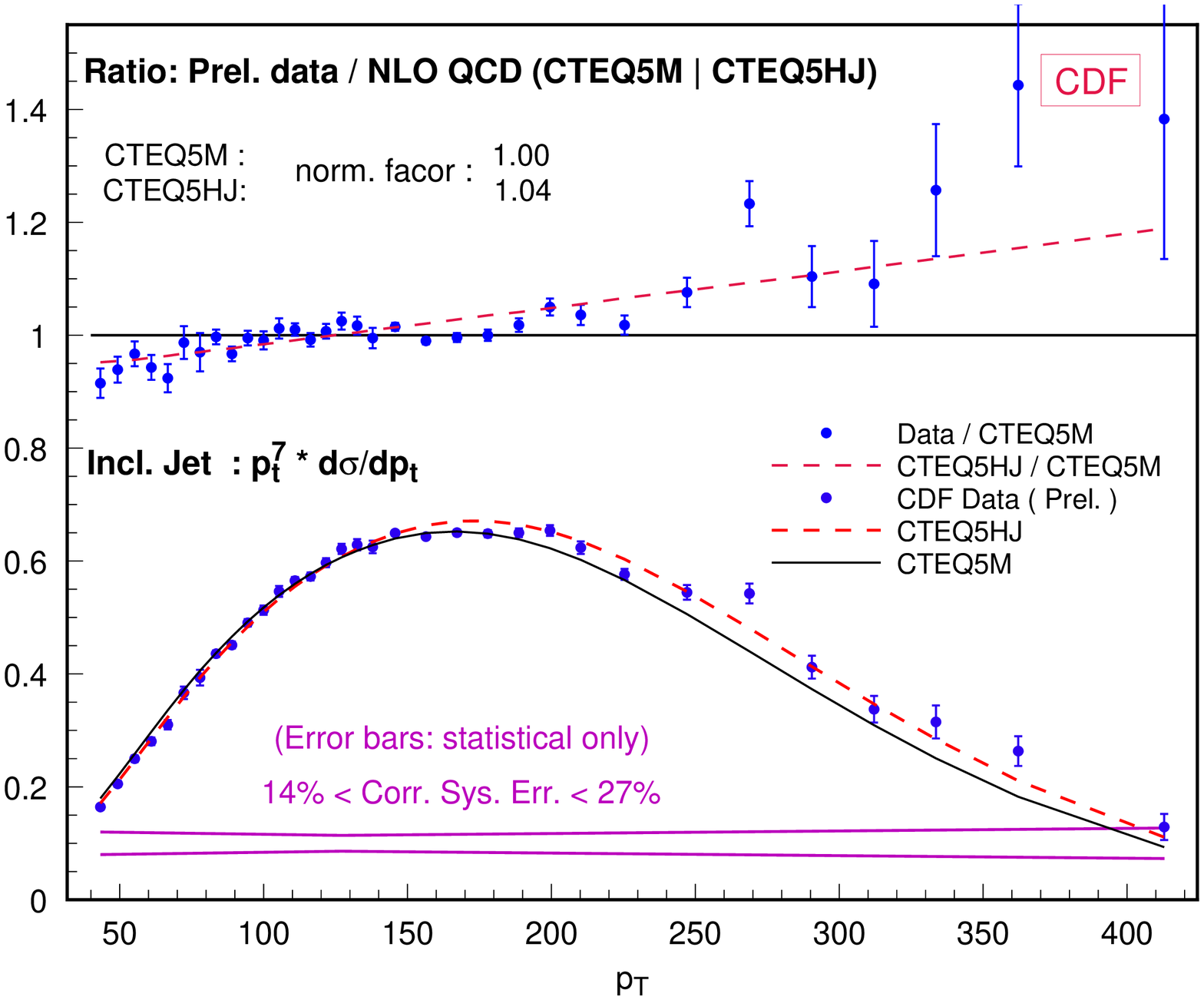}&
\includegraphics*[height=5.7cm]{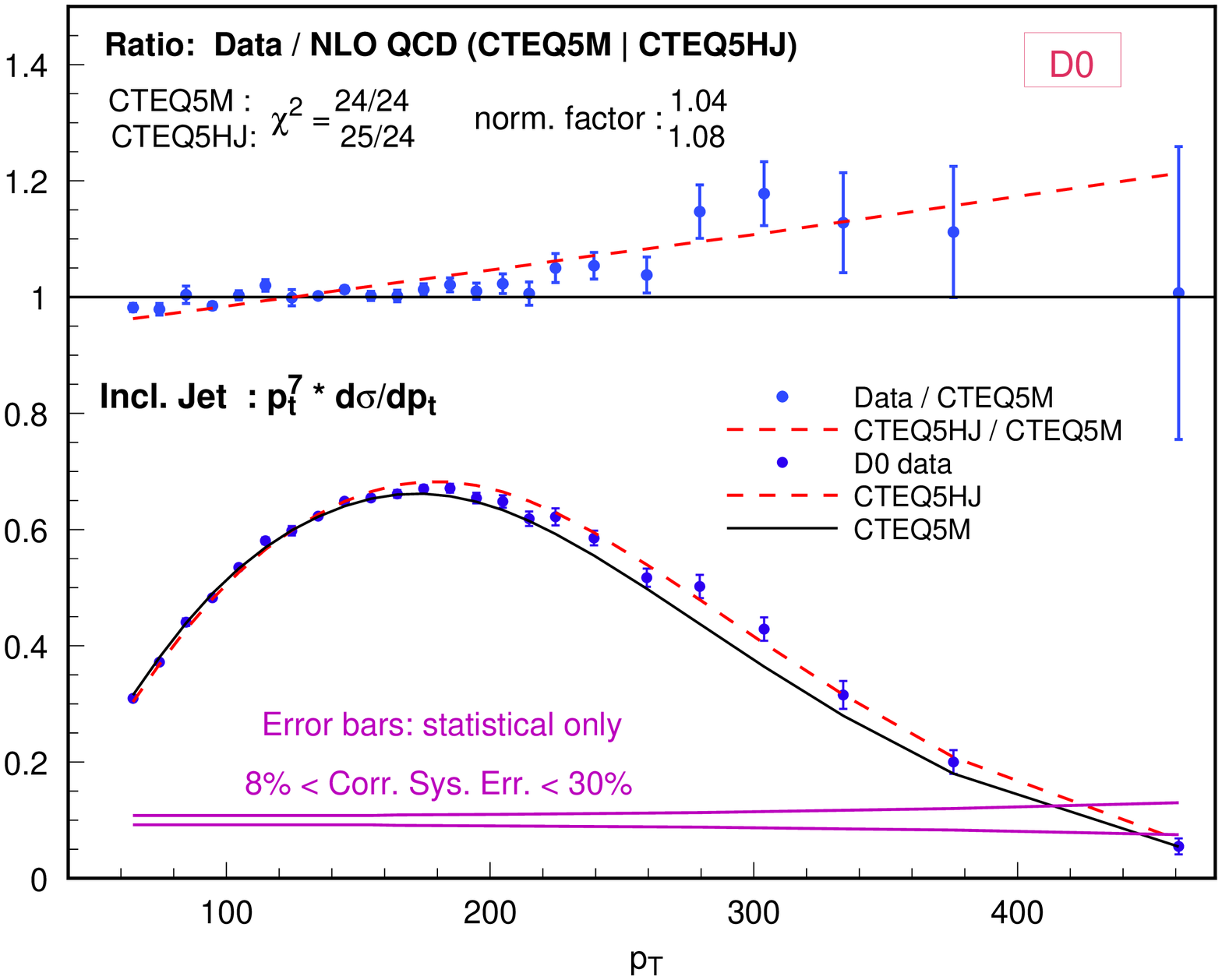}\\
\end{tabular}
\caption{Inclusive jet cross sections measured by CDF (left)
and D\O\ (right), compared to the NLO QCD prediction using CTEQ5
parton distributions. The upper data points show {\it (data--theory)/theory},
while the lower points are the measured and predicted
cross sections (approximately linearized
though multiplication by $E_T^7$). 
\label{fig:cteq5}}
\end{center}
\end{figure}

In order to compare with CDF, D\O\ carried out an analysis in exactly
the same rapidity interval ($0.1 < |\eta| < 0.7$).  The results\cite{annrev}
are shown in Fig.\ref{fig:annrev}.  Firstly we note that there is no 
actual discrepancy between the datasets. Secondly, for this plot the 
theoretical prediction was made using the CTEQ4HJ parton distribution,
which has been adjusted to give an increased gluon density at large $x$ while
not violating any experimental constraints (except perhaps fixed target photon
production data, which in any case
require big corrections before they can be compared to QCD, as we shall
see later).  The result of this increased gluon content is improved
agreement especially with the CDF data points.  The situation with
the latest CTEQ5M and CTEQ5HJ parton distributions is shown in 
Fig.~\ref{fig:cteq5}, and again, the enhanced gluon content in CTEQ5HJ
brings the predicted cross section closer to the CDF data.

What then have we learned from this issue?  In my opinion, whether the
CDF data show a real excess above QCD, or just a ``visual excess,''
depends critically on understanding the systematic errors and their
correlations as a function of $E_T$.  Whether nature has actually exploited
the freedom to enhance gluon distributions at large $x$ will only be clear 
with the addition of more data --- the factor of 20 increase in luminosity
in the first part of Run~2 will extend the reach by
70--100~GeV in $E_T$ and should therefore make
the asymptotic high-$x$ 
behavior clearer.  Whatever the Run~2 data show, this has
been a useful lesson; it has reminded us all that parton distributions
have uncertainties, whether made explicit or not, and that a full 
understanding of experimental systematics {\it and their correlations} is needed
to understand whether experiments and theory agree or disagree.   

\begin{figure}[p]
\begin{center}
\begin{tabular}{cc}
\includegraphics*[bb=100 200 525 600,height=6cm]{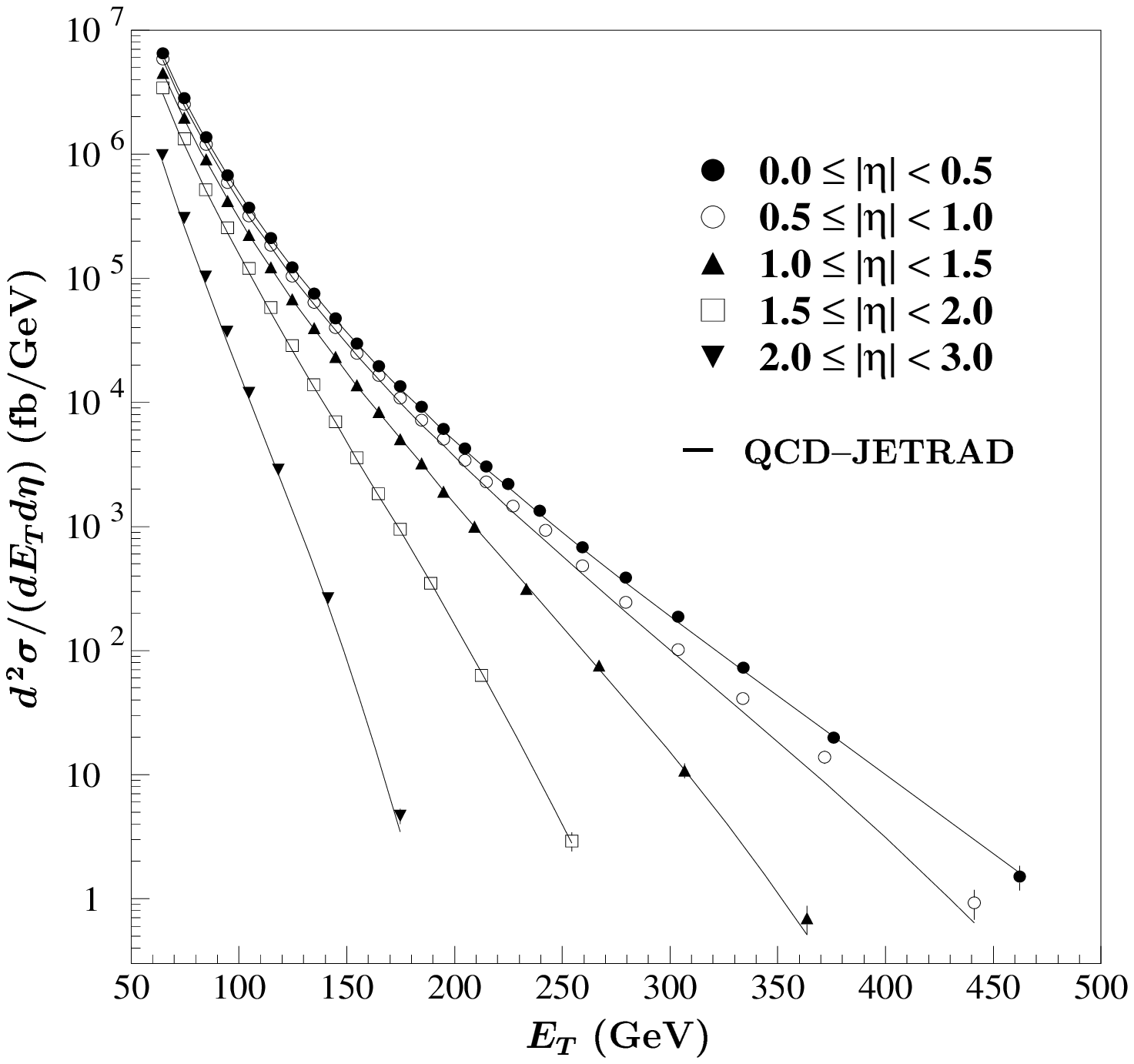}&
\includegraphics*[bb=30 160 525 655,height=6cm]{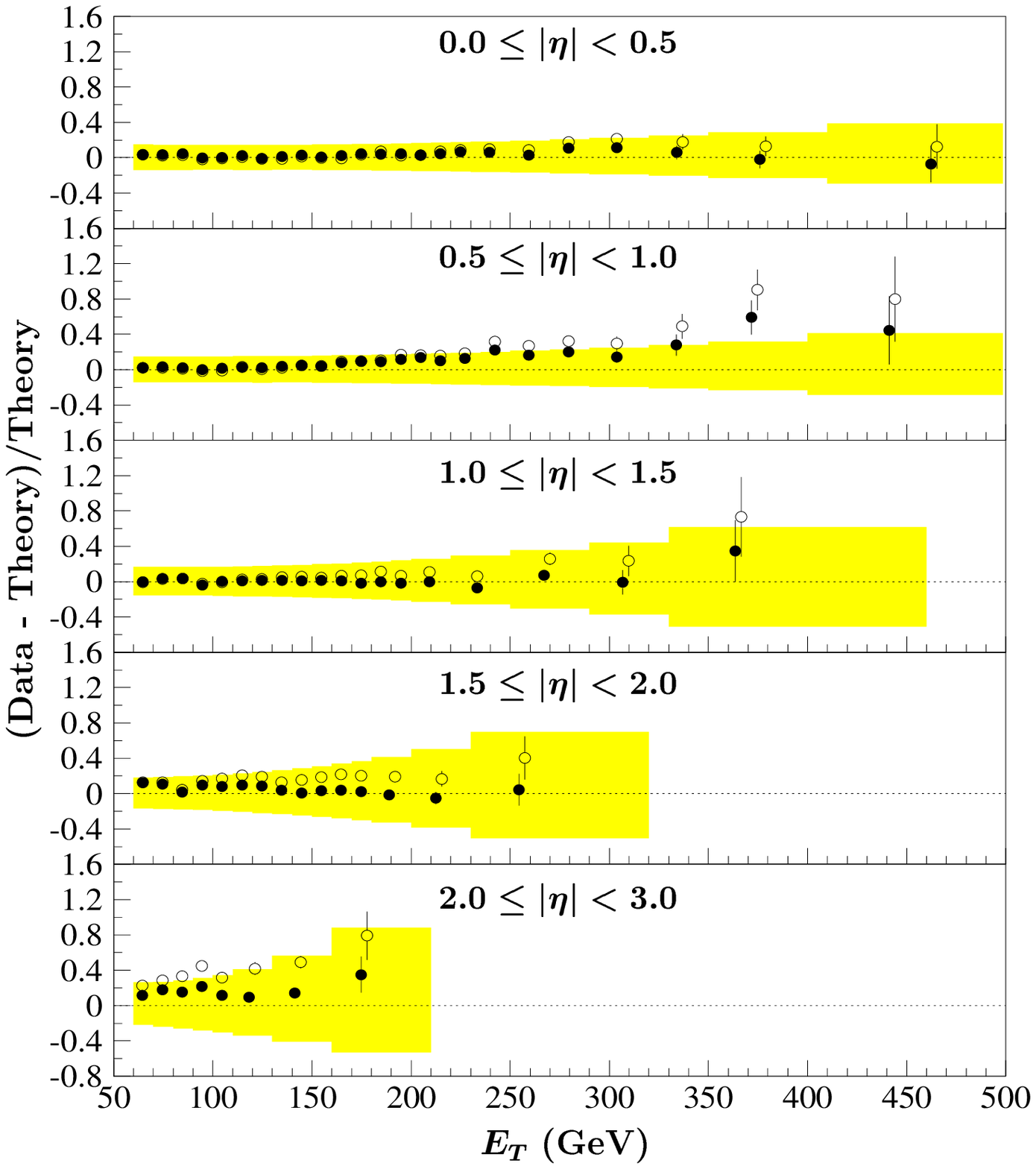}\\
\end{tabular}
\caption{Inclusive jet cross sections measured up to $|\eta|=3$ 
by D\O\protect\cite{d0fwdjets},
compared to the NLO QCD prediction (using the JETRAD Monte Carlo).
In the left hand plot the prediction uses CTEQ4M; in the right hand
plot the solid points use CTEQ4HJ while the open points use CTEQ4M.  
\label{fig:fwdjets}}
\vspace{2cm}
\begin{tabular}{cc}
\includegraphics*[height=6.5cm]{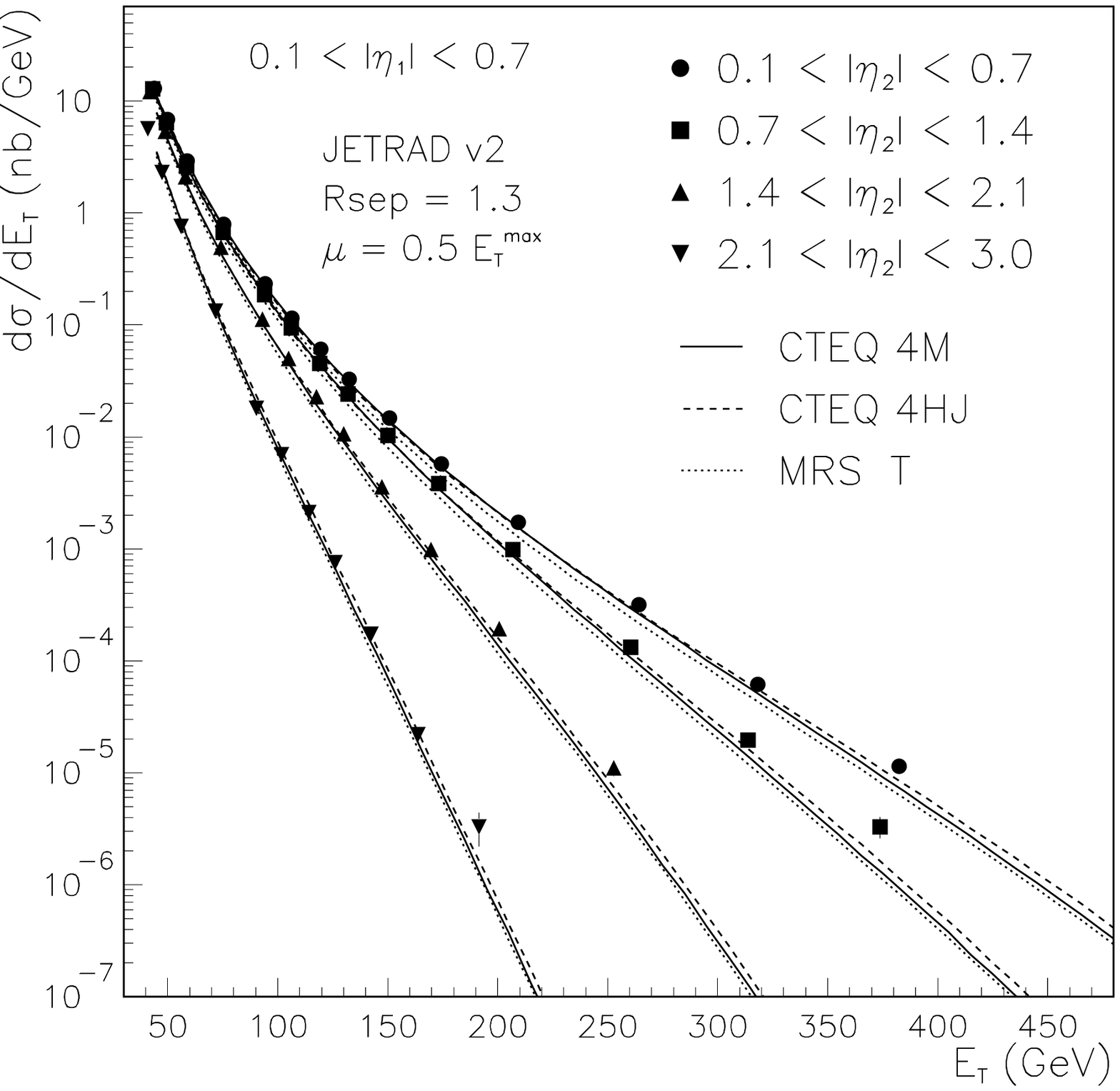}&
\includegraphics*[height=6.5cm]{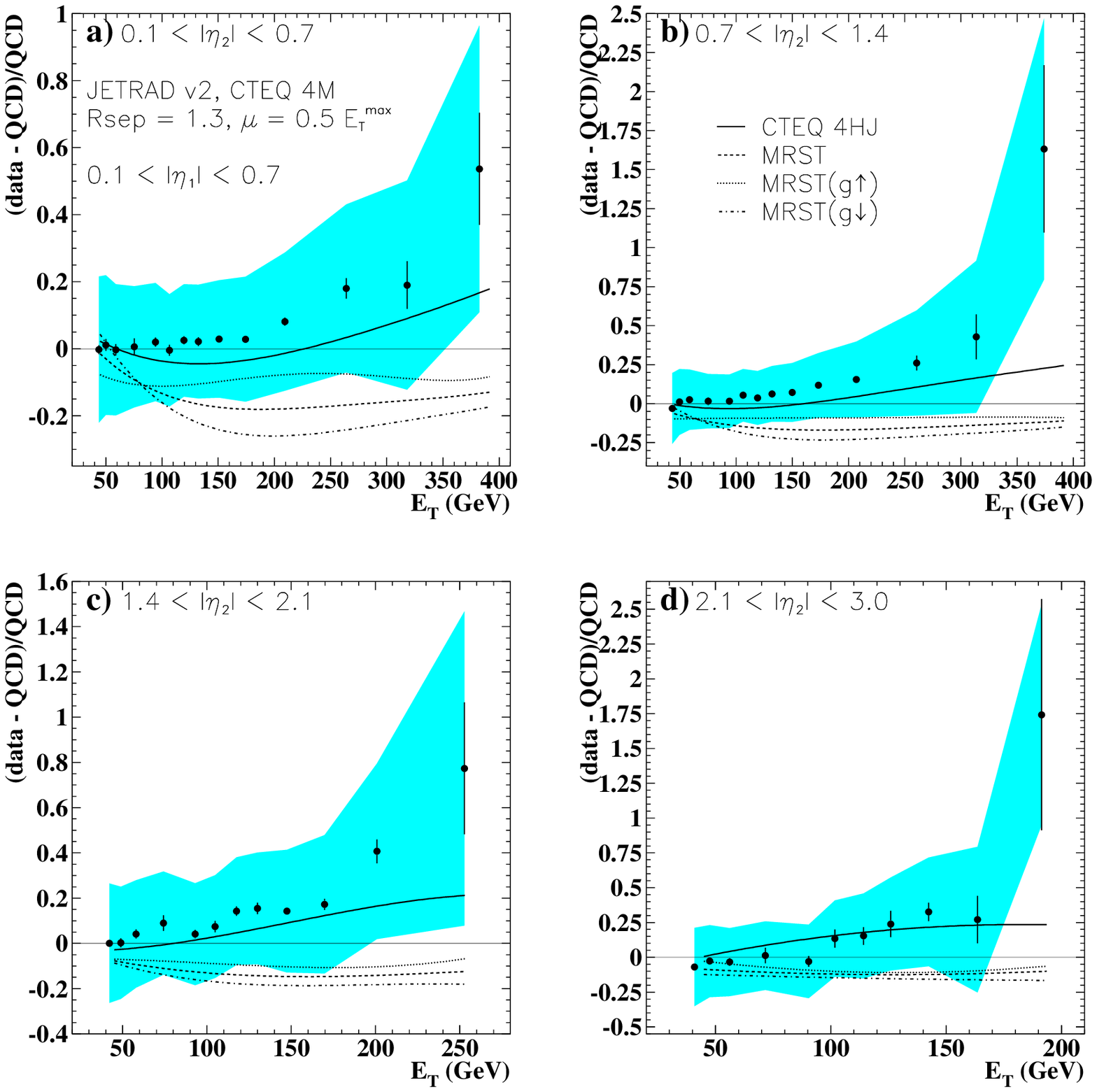}\\
\end{tabular}
\caption{Dijet cross sections measured
by CDF\protect\cite{cdftripdiff} for events with one central
jet $0.1<|\eta_1|<0.7$ and one jet allowed forward;
left, as a function of the central jet $E_T$ for various bins
of $|\eta_2|$, and right, normalized to the NLO QCD prediction
(from JETRAD).}
\label{fig:cdftripdiff}
\end{center}
\end{figure}

D\O\cite{d0fwdjets} 
have extended their measurement of inclusive jet cross sections into
the forward region.  Figure~\ref{fig:fwdjets} shows the measured cross
sections up to  $|\eta|=3$.  They are in good agreement with NLO
QCD over the whole range of pseudorapidity and transverse energy; in
fact both CTEQ4M and CTEQ4HJ parton distruibutions yield a good 
$\chi^2$. 

Both Tevatron experiments have also studied dijet final states.
CDF\cite{cdftripdiff} has presented cross sections for processes
with one central jet ($0.1 < |\eta_1| < 0.7$) and one jet allowed
forward ($|\eta_2|$ up to 3.0).  In Fig.~\ref{fig:cdftripdiff} these
are compared with the NLO QCD prediction as a function of the 
central jet's transverse energy ($E_{T1}$). 
The data show an excess above the theory for large $E_{T1}$,
just as seen in the inclusive cross section; but since these events
are common to both samples, this is not surprising.

D\O\ have measured\cite{d0fwdjets} 
the cross sections for dijet production with both 
same-side ($\eta_1 \approx \eta_2$)
and opposite-side ($\eta_1 \approx - \eta_2$) topologies, for four
bins of $|\eta|$ up to 2.0.  The results are all in good agreement
with the NLO QCD prediction.

\begin{figure}[tb]
\begin{center}
\includegraphics*[bb=60 200 525 655,height=8cm]{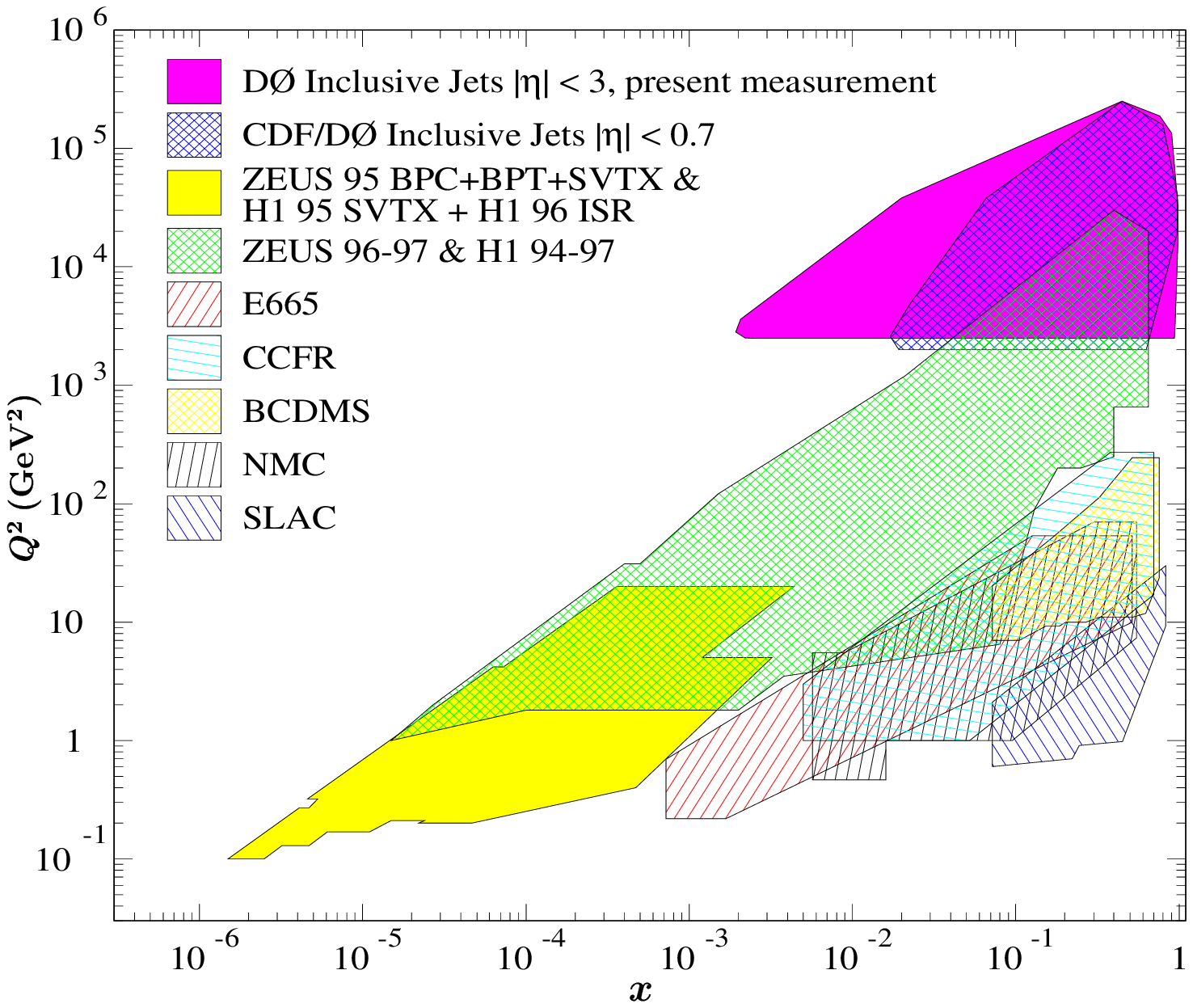}
\caption{Phase space in $x$ and $Q^2$ probed by various
experiments, showing the potential of central and forward jet
production at the Tevatron to constrain parton distributions
in regions complementary to deep inelastic scattering.
Taken from \protect\cite{d0fwdjets}.
\label{fig:xqsquared}}
\end{center}
\end{figure}

All of these central, forward and dijet cross section 
measurements should really be used as
input to the parton distribution fitting ``industry''.  
Figure~\ref{fig:xqsquared} shows where the Tevatron data lie
on the plane of $x$ and $Q^2$, indicating their complementarity
to the fixed target and HERA deep-inelastic data.  The apex of
the Tevatron phase space is set by the highest $Q^2$ event
observed in Run~1, a spectacular dijet seen in D\O\ with a jet-jet
invariant mass
of 1.2~TeV, $Q^2 = 2.2 \times 10^5$GeV$^2$, and $x_1 = x_2 = 0.66$.

\subsection {Extraction of $\alpha_s$}

\begin{figure}[p]
\begin{center}
\includegraphics*[height=8cm]{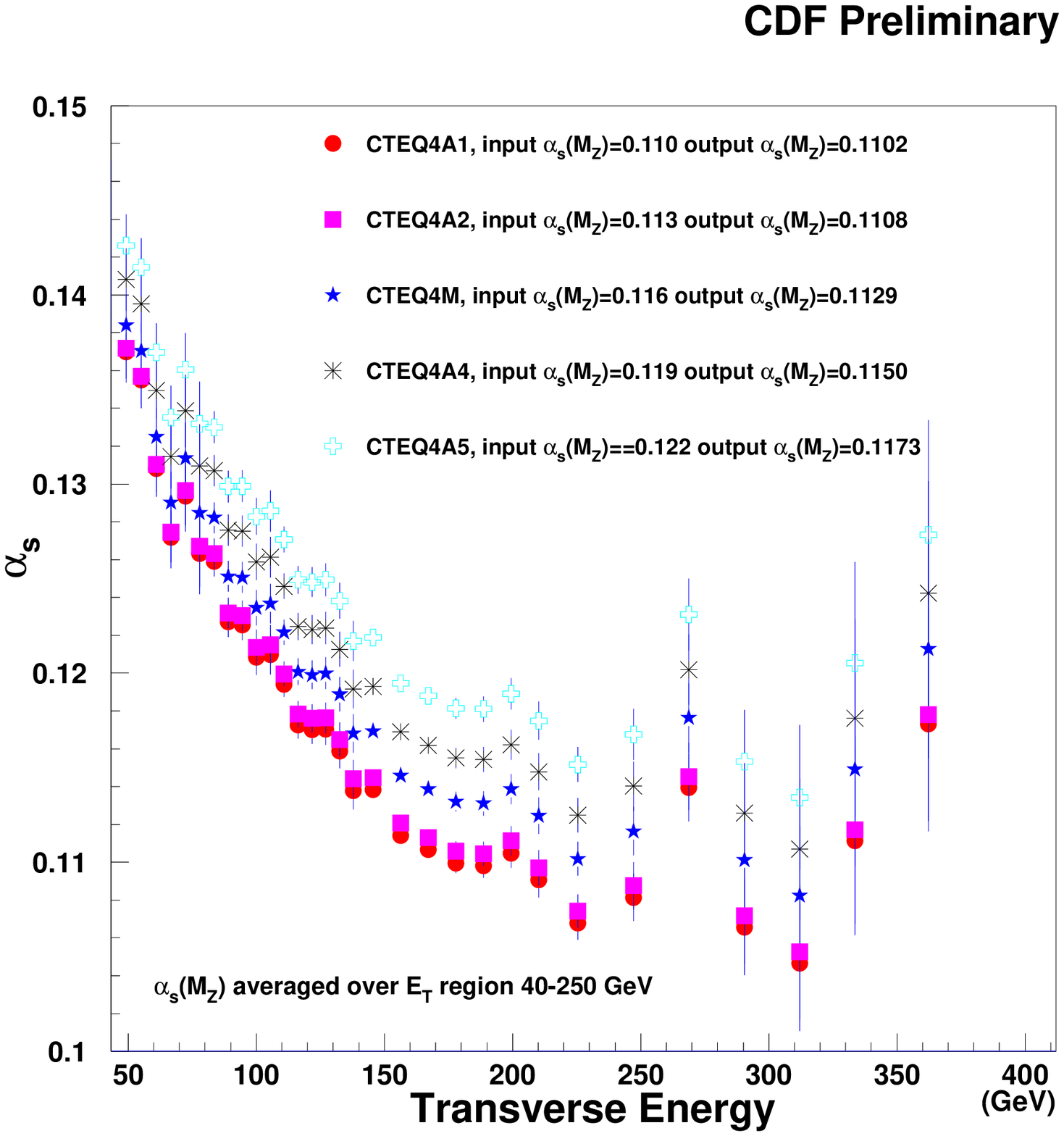}
\caption{Value of $\alpha_s$ as a function of scale (jet transverse 
energy) inferred by CDF from the inclusive
jet cross section using the CTEQ4A
series of parton distributions\protect\cite{cdfalphas}.}
\label{fig:cdfalphas}
\vspace{2cm}
\begin{tabular}{cc}
\includegraphics*[height=6.6cm]{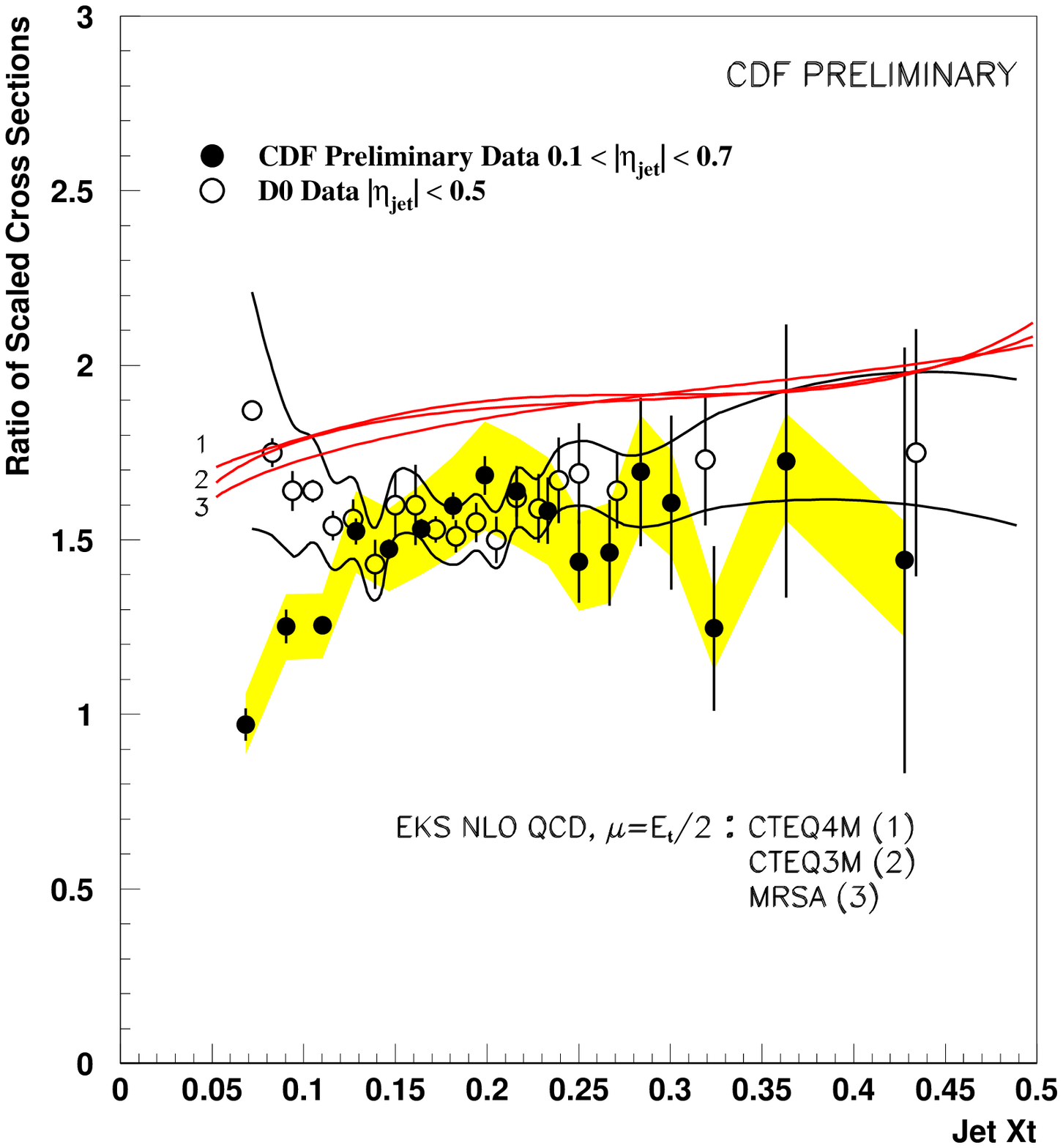}&
\includegraphics*[height=6cm]{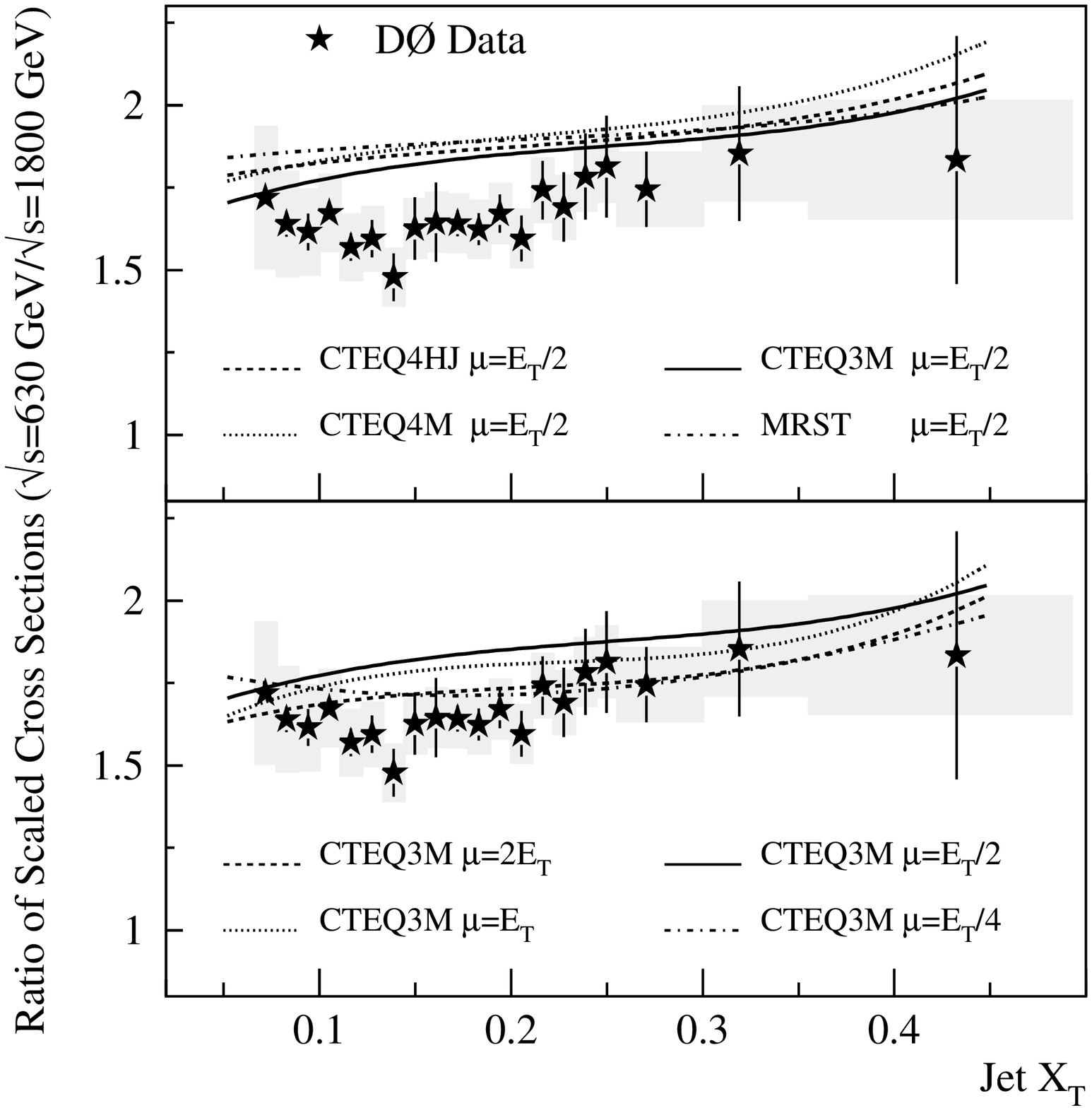}\\
\end{tabular}
\caption{Scaled ratios of jet cross sections at $\sqrt{s}=630~GeV$ 
to $\sqrt{s}=1800~GeV$, as a function of $x_T = 2E_T/\sqrt{s}$,
as measured by CDF\protect\cite{cdf630} and D\O\protect\cite{d0630} 
and as predicted by NLO QCD.
\label{fig:630}}
\end{center}
\end{figure}

CDF have carried out an interesting study with the aim
of extracting $\alpha_s$ from 
the inclusive jet cross section\cite{cdfalphas}; at NLO,
the calculated cross section depends on $\alpha_s$ with a
coefficient which is predicted by JETRAD.
The result, 
$\alpha_s(m_Z) = 0.113^{+0.008}_{-0.009}$, is consistent with the world
average, and $\alpha_s$ shows a nice evolution with scale (given by the
jet transverse energy), as shown in Fig.~\ref{fig:cdfalphas}.  
However the figure also shows that the measurement suffers from a large, 
and hard to quantify, sensitivity to the parton distributions,
especially to the value of $\alpha_s$ assumed therein.  
At this time I think it must be characterized as a nice test of QCD and not
really as a measurement of $\alpha_s$.

\subsection{Cross section ratio 630~GeV/1800~GeV}

Both CDF\cite{cdf630} and D\O\cite{d0630} have exploited
a short period of data taking at reduced center of mass
energy towards the end of Run~1, to measure the
ratio of scale invariant jet cross sections,
$E_T^3 d^2\sigma/dE_Td\eta$ at $\sqrt{s}=1800$ and 630~GeV.
This ratio, as a function
of scaled jet transverse energy $x_T = 2E_T/\sqrt{s}$, is shown in
Fig.\ref{fig:630}.  The ratio is expected to be a rather straightforward
quantity to measure and to calculate --- it would be exactly 1 in the
pure parton model.  
Unfortunately the two experiments are not obviously
consistent with each other (especially at low $x_T$) nor with NLO QCD
(at any $x_T$).  At least two explanations have been suggested for
the discrepancy.
Firstly, different renormalization scales could be used for the
theoretical calculations at the two energies.  While allowed, this
seems unappealing.\footnote{Glover has suggested that such
a procedure is in fact natural when a scaling variable
like $x_T$ is used; because $x_T$ differs by a factor of about 
three between the two center of mass energies for a given $E_T$,
a factor of three difference in the renormalization scales is appropriate.} 
An alternative explanation is offered by Mangano\cite{mangano}, 
who notes that a shift of a few GeV
in energy between parton and particle level jets  
would bring the data in line with the prediction.  
Such a shift might arise from non-perturbative effects
such as losses outside the jet cone, underlying event energy, and 
intrinsic transverse momentum of the incoming partons; the shifts
would likely be jet algorithm-dependent,
and the two experiments might even obtain
different results depending on how the jet energy scale
corrections were done (based on
data or Monte Carlo, for example).  It seems that more work, both
theoretical and exerimental, is needed before this question can be resolved.

\subsection{Ratio of 3-jet/2-jet Events}

\begin{figure}[tb]
\begin{center}
\begin{tabular}{cc}
\includegraphics*[height=6cm]{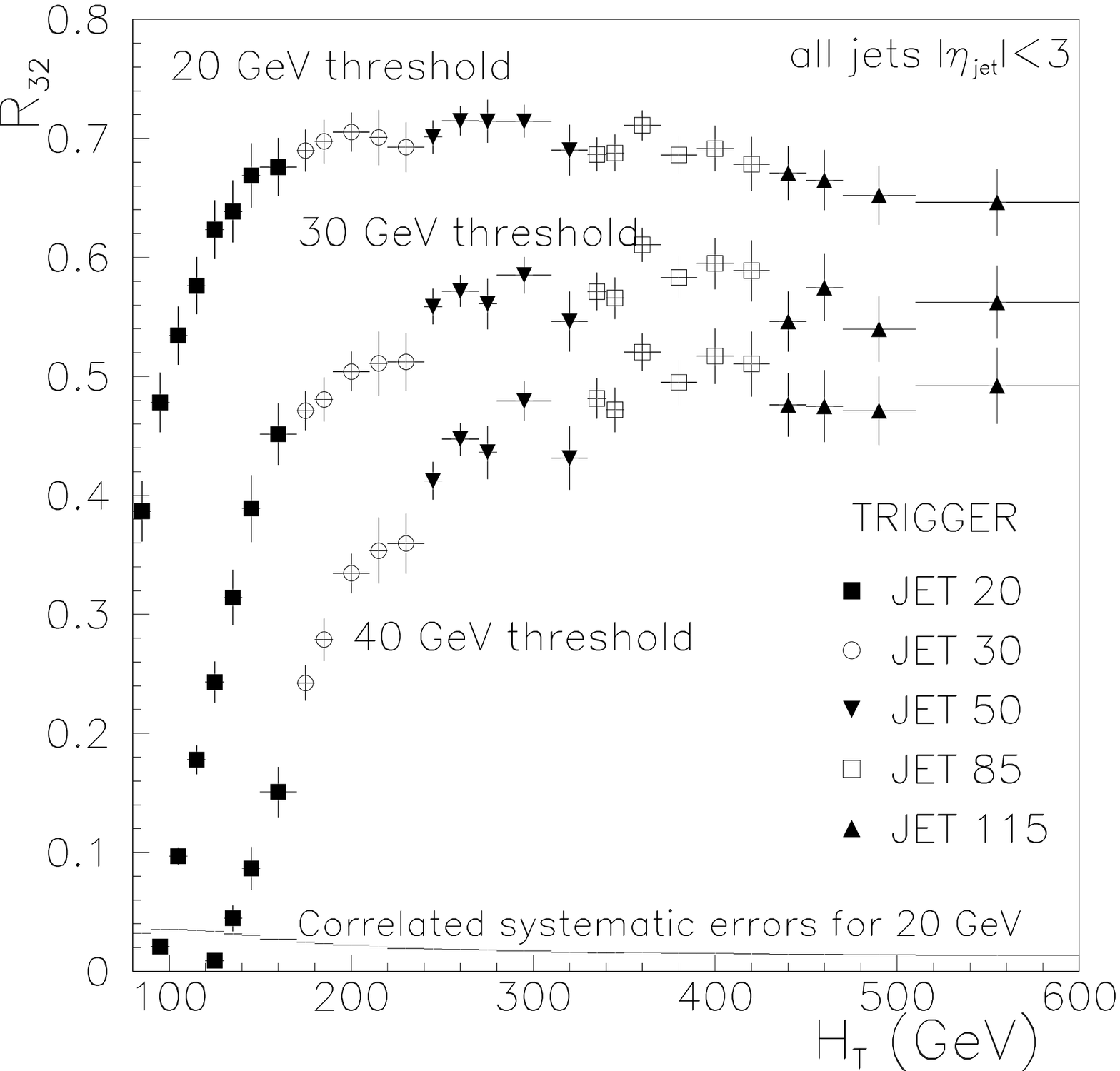}&
\includegraphics*[height=6cm]{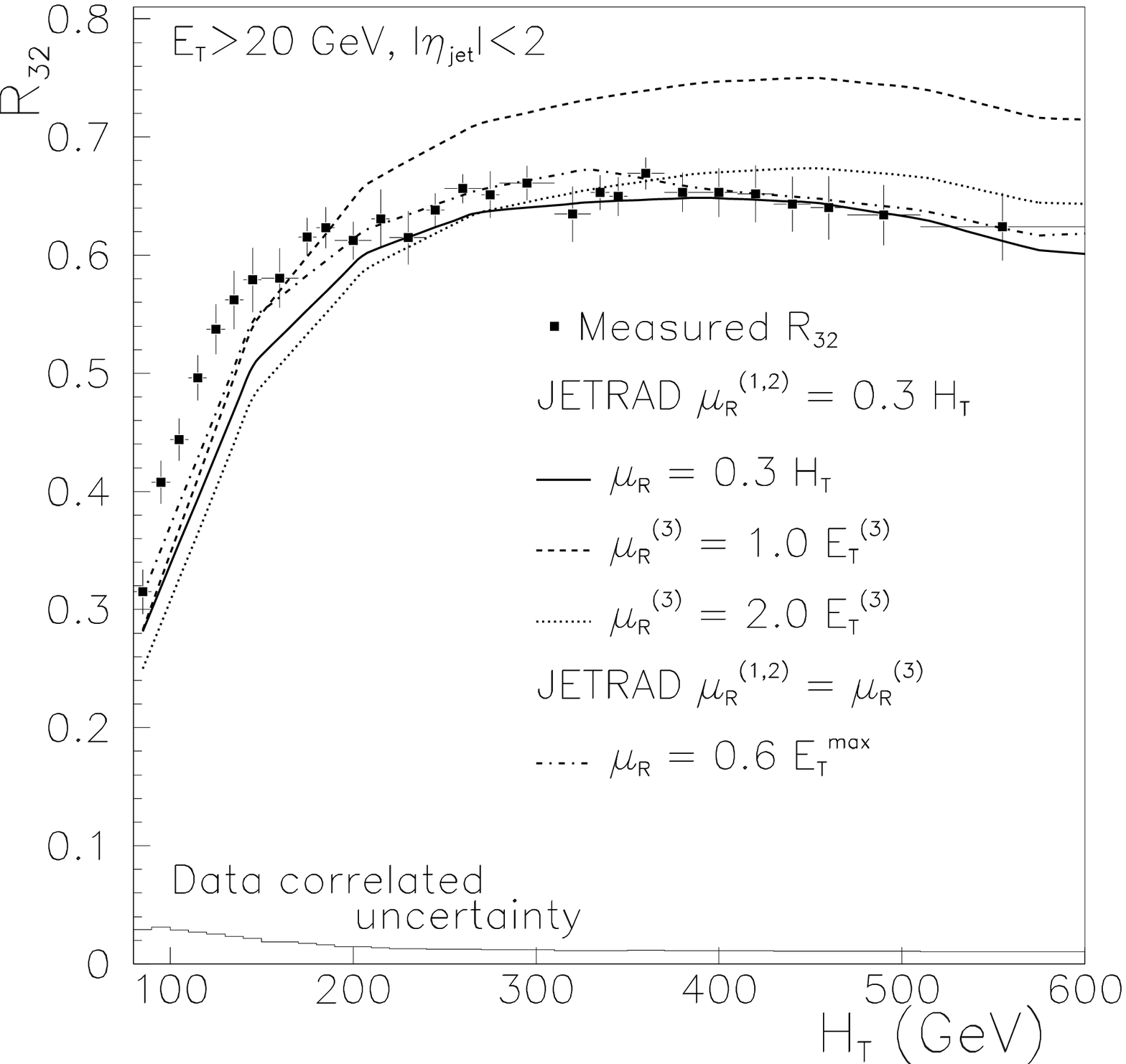}\\
\end{tabular}
\caption{Ratio of events with three or more jets to those
with two or more jets, $R_{32}$, measured as a function
of $H_T = \Sigma E_T^{\rm jets}$, for various jet thresholds (left)
and compared with JETRAD (right).\protect\cite{d0gallas} 
\label{fig:r32}}
\end{center}
\end{figure}

D\O\cite{d0gallas} have measured $R_{32}$, the ratio of
events with $\geq 3$~jets to those with $\geq 2$~jets, as
a function of $H_T = \Sigma E_T^{\rm jets}$, for various third jet
thresholds.  This ratio (Fig.~\ref{fig:r32}) is
surprisingly large:  two thirds of high-$E_T$ jet events have
a third jet with $E_T > 20$~GeV and about half have a third
jet above 40~GeV.  It is interesting to ask if this ratio
can be predicted by QCD.  The answer is yes, reasonably well,
even by JETRAD (which of course
is a leading order calculation of $R_{32}$).
D\O\ have also attempted to extract information on the optimal
renormalization scale for the emission of the third jet: 
should it be the same scale as the leading jets, or should
the third jet emission be treated as part of a parton shower
with an ``evolving'' scale related to the third jet's $E_T$?
(A specially modified version of JETRAD was used for this study).
They find that a scale tied to the first two jets is better
than one related to the third jet $E_T$.  Whether this tells 
us much about nature or merely about JETRAD I don't know, but
it's interesting given the widespread use of the parton shower
approximation to generate additional jets in HERWIG and PYTHIA.

\subsection{Jet Structure and Quark/Gluon Separation}

\begin{figure}[tb]
\begin{center}
\includegraphics*[height=5cm]{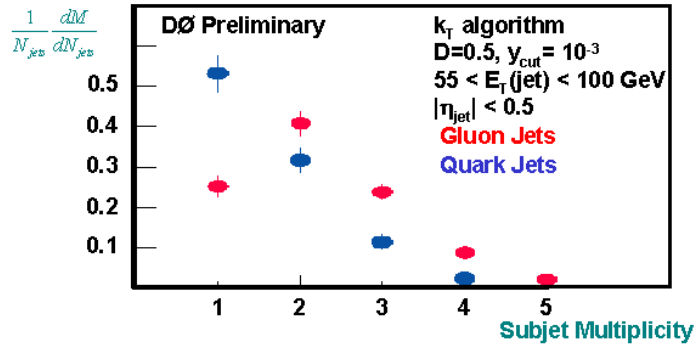}
\caption{Subjet multiplicities measured by D\O\ using a 
$k_T$ algorithm to find clusters within jets; distributions 
for quark and gluon
jets are inferred using $\sqrt{s}=1800$~GeV and
$\sqrt{s}=630$~GeV data\protect\cite{d0snihur}.
\label{fig:subjets}}
\end{center}
\end{figure}

All the results presented so far have used a cone jet finder.
By running a $k_T$ jet finder inside previously identified jets,
one can count the number of ``subjets'' or energy clusters.  
Doing this (rather than, for example, counting charged tracks)
allows the {\it coarse} jet structure corresponding to the 
initial, perturbative part of fragmentation to be studied.
D\O\cite{d0snihur} have made such a measurement and, by comparing
jets of the same $E_T$ and $\eta$ recorded at $\sqrt{s}=1800$ and 630~GeV,
have inferred the composition of pure quark and gluon jets.
The extracted subjet multiplicity $M$ for the two species is shown
in Fig.\ref{fig:subjets}.  The ratio of $M-1$ for the two cases,
which might naively be expected to equal the ratio of gluon and
quark colour charges,
is found to be $1.91\pm0.04$, compared with $1.86\pm0.04$ from HERWIG.
This is very encouraging and might even suggest that we have glimpsed
the holy grail of quark-gluon jet separation.  The true test, however,
remains the use of the subject multiplicity as a discriminant in
an analysis like the search for $t \overline t \to 6$~jets.  Such a
test will probably have to wait for Run~2.

\section{Weak Boson production}

\begin{figure}[tb]
\begin{center}
\includegraphics*[height=8cm]{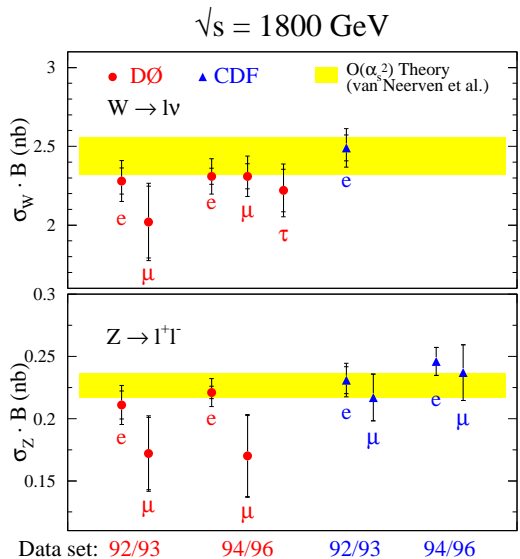}
\caption{Vector boson production cross sections measured at the
Tevatron by CDF and D\O, compared to the NNLO QCD prediction.}
\label{fig:wzxsec}
\end{center}
\end{figure}

Next-to-next-to leading order
(order $\alpha \cdot \alpha_s^2$) predictions exist for the $W$ and $Z$
production cross sections times decay branching ratios into leptons.
The experimental values from CDF and D\O\ (Fig.~\ref{fig:wzxsec})
are in excellent agreement with these predictions,
both for electrons and muons.  In fact, the careful
reader will note that the CDF cross sections are a 
few percent higher than those from D\O; this is consistent with the
fact that CDF use a luminosity normalization which is 6.2\% higher
than D\O's (the two experiments assume different total $p\overline p$
inelastic cross sections).  It is therefore tempting to conclude that
the $W/Z$ cross sections are the better known quantity, and
indeed it has been seriously proposed to use $\sigma_W$ as the
absolute luminosity normalization basis in Run~2.  Walter Giele's
contribution in these proceedings contains some more discussion of the
systematics associated with such an approach.

\begin{figure}[p]
\begin{center}
\includegraphics*[height=6.5cm]{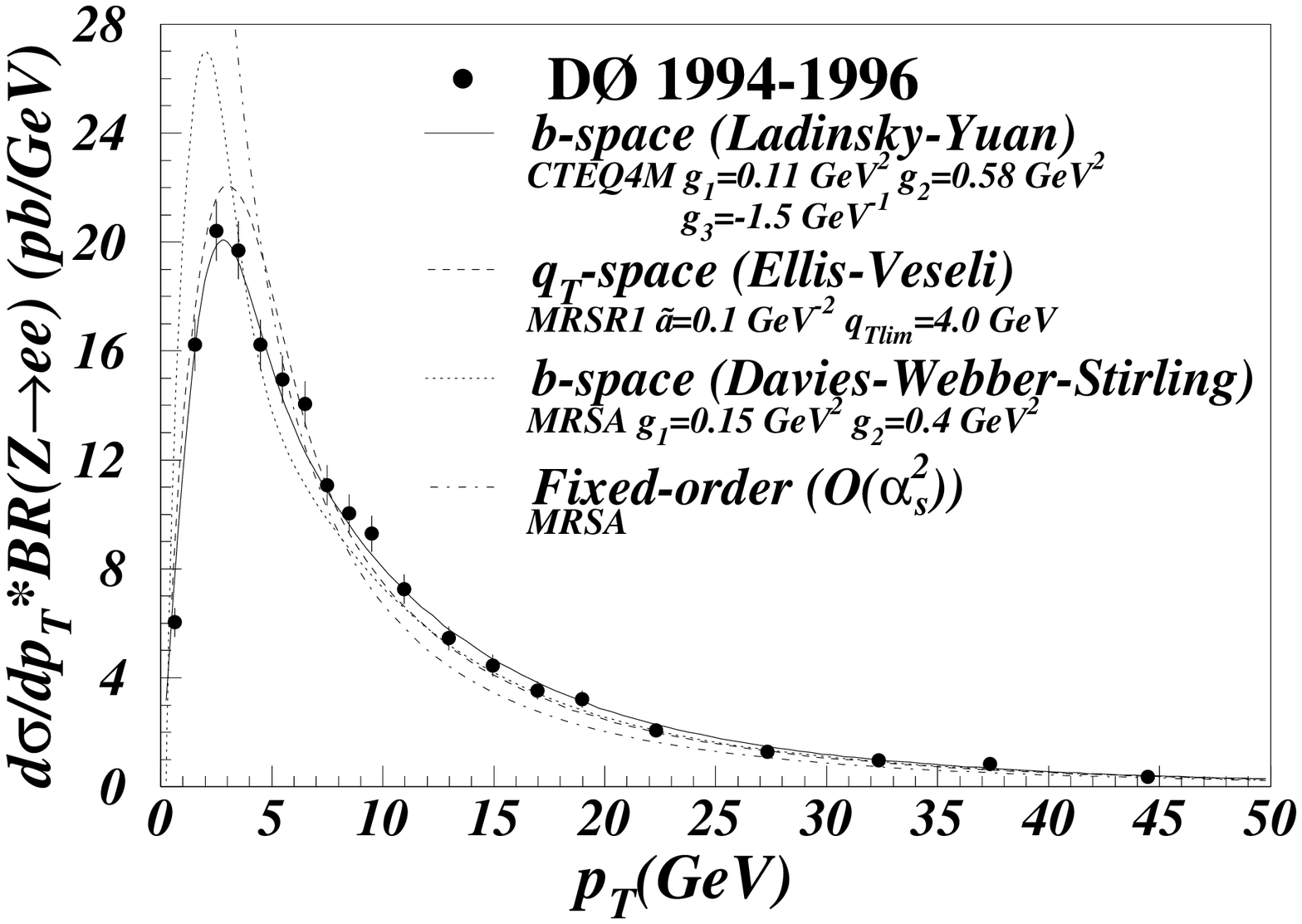}
\begin{tabular}{cc}
\includegraphics*[height=4cm]{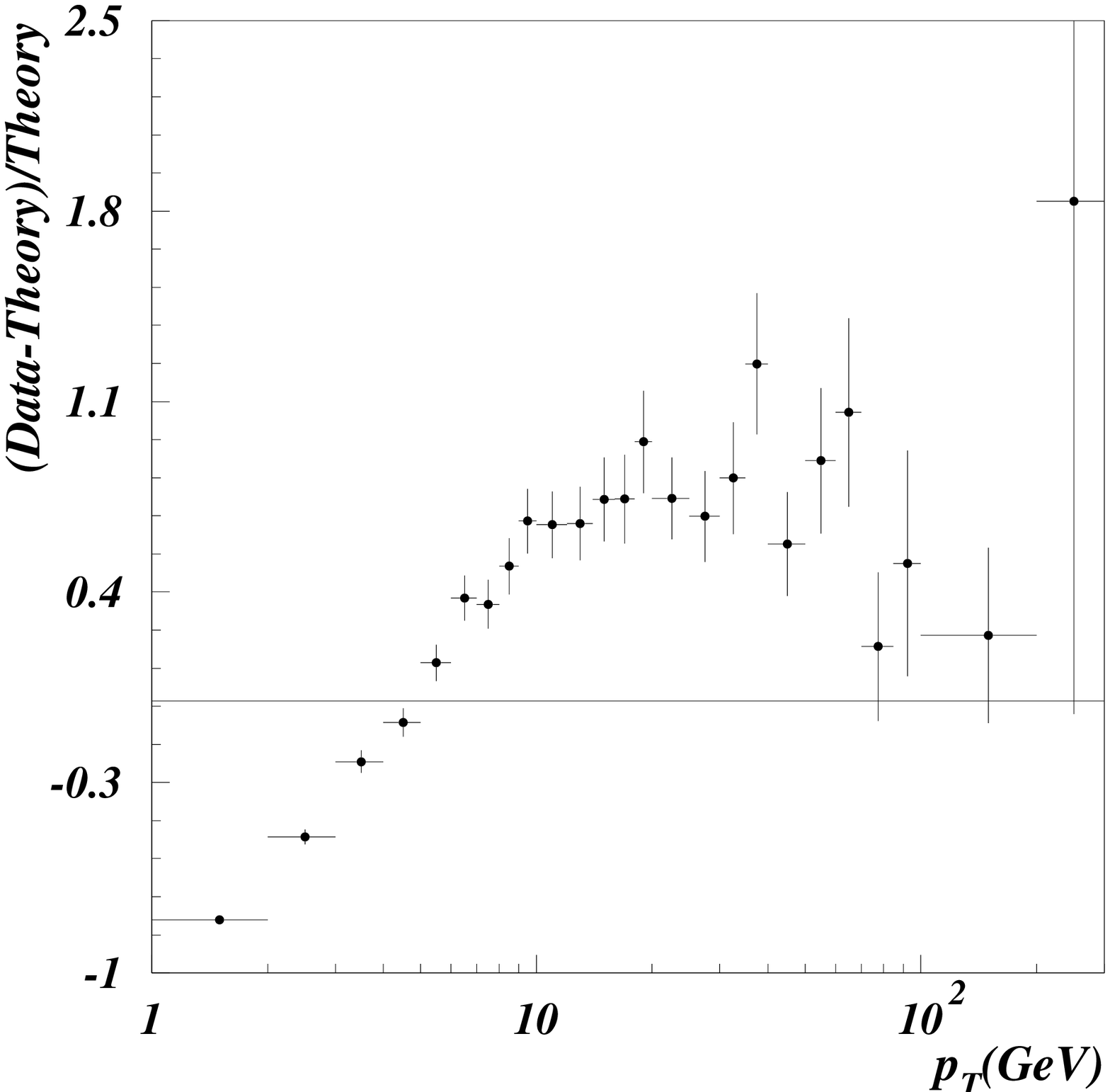} &
\includegraphics*[height=4cm]{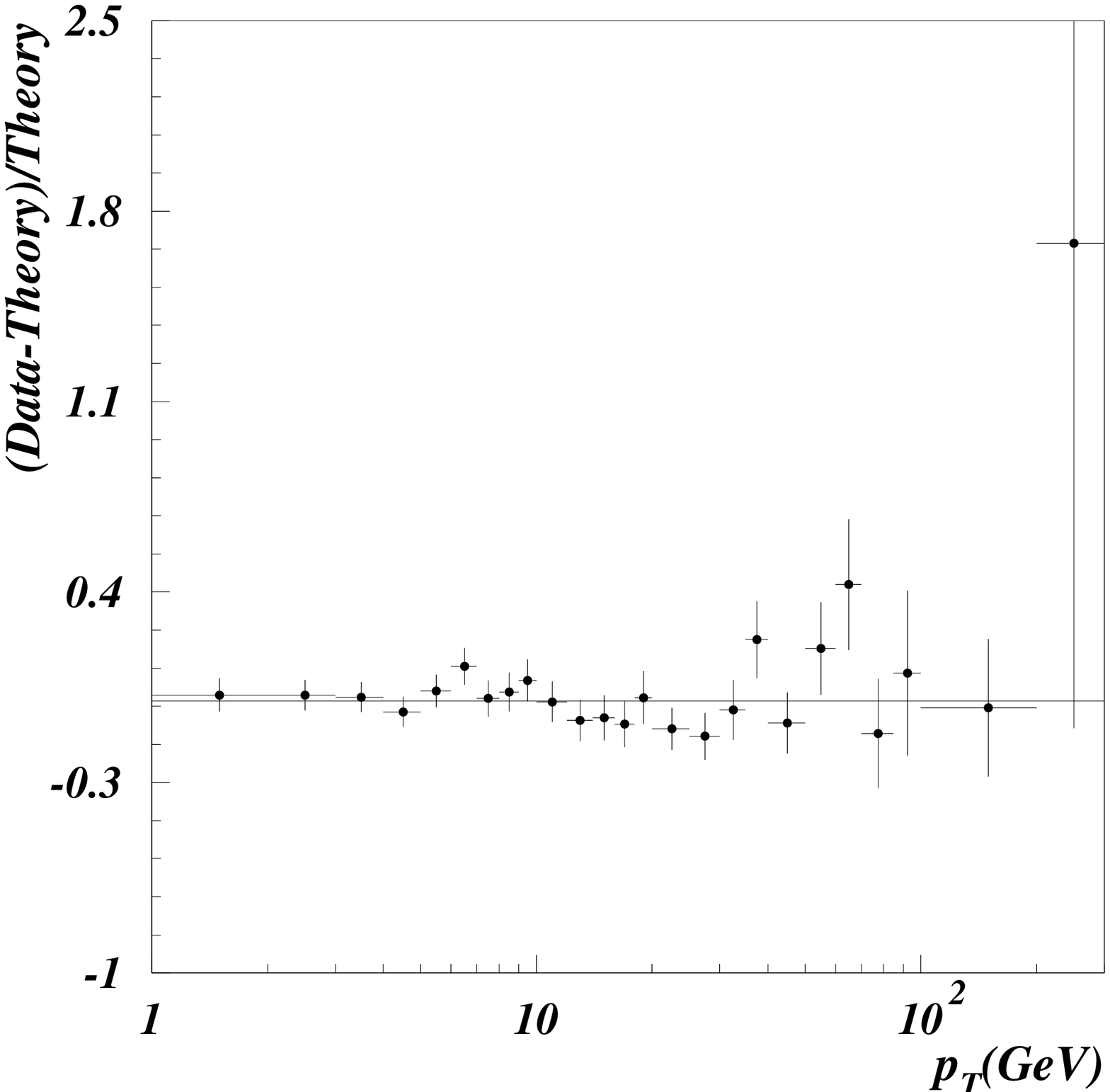} \\
\end{tabular}
\caption{Transverse momentum distribution of the $Z$, as measured by
D\O\protect\cite{d0ptz}. 
The upper plot shows the data and various calculations.  The
lower left shows the data normalized to the fixed-order QCD prediction and
the lower right shows the data normalized to the resummed calculation
of Ladinsky and Yuan\protect\cite{ladinsky}.    
\label{fig:ptz}}

\includegraphics*[height=6.5cm]{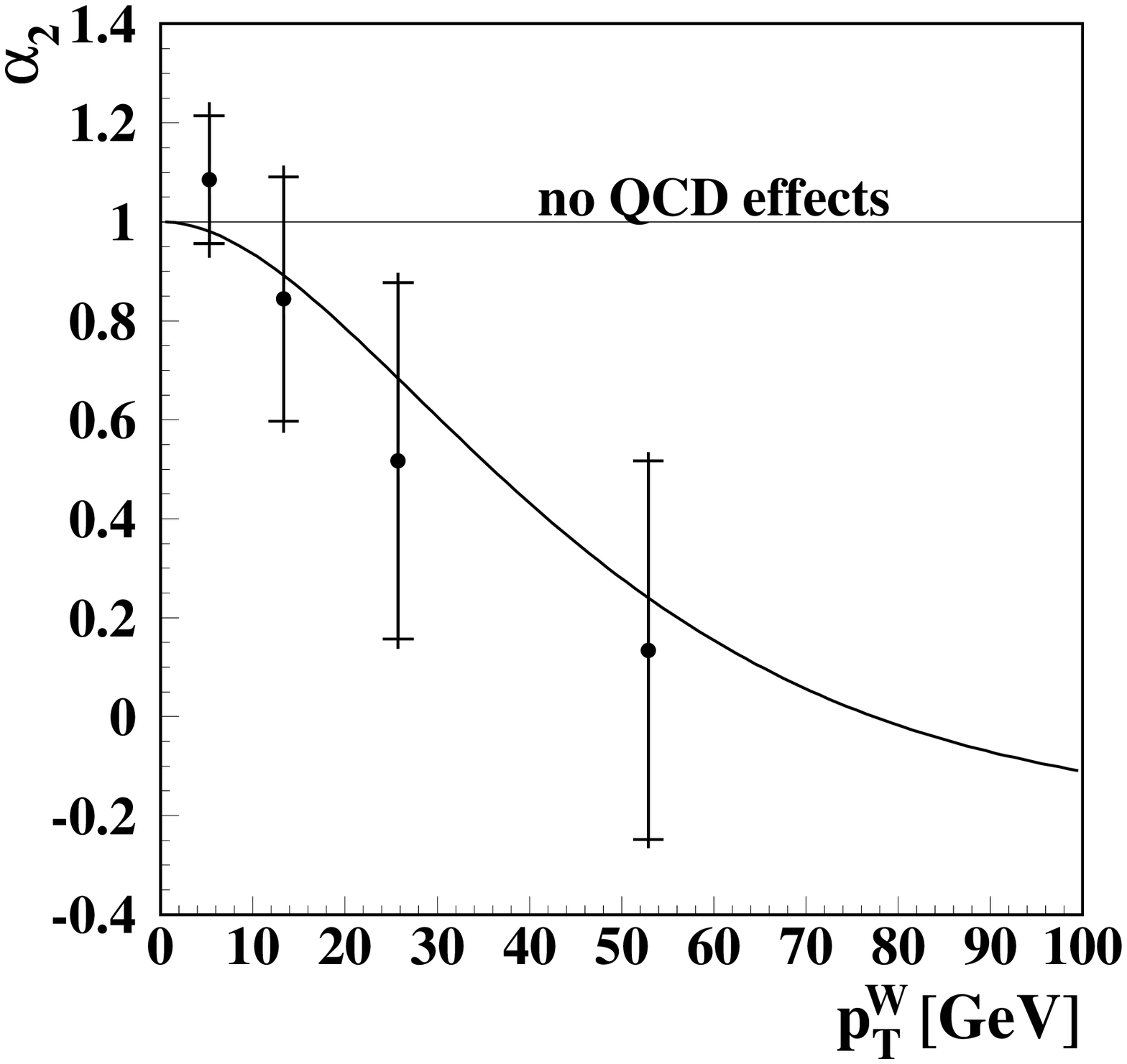}
\caption{Extracted value \protect\cite{georg}    
of $\alpha_2$ (the coefficient of 
$\cos^2\theta^*$ in the angular distribution of electrons
from $W$ decay) as a function of $p_T^W$ .  The measurement is in
good agreement with QCD.
\label{fig:georg}}
\end{center}
\end{figure}

\subsection{$Z$ Transverse Momentum}

As well as increasing the total cross section, the QCD predictions
change the transverse momentum distribution of the produced
boson.  The most straightforward measurement is for the $Z$ since
it can be directly reconstructed from two decay leptons.
Figure~\ref{fig:ptz} show recent
D\O\ results on the transverse momentum distribution
of the $Z$ boson\cite{d0ptz} compared with a variety of QCD 
predictions.  Clearly the
fixed-order NLO QCD is not a good match for the data, while the
resummed formalism of Ladinsky and Yuan\cite{ladinsky} 
fits rather well.  This approach uses fixed-order QCD at
high $p_T^Z$ matched to a resummation of the large
logarithms of $m_Z^2/p_T^2$ at low $p_T^Z$. 
The resummed calculations
always include some nonperturbative parameters that must be
extracted from the data, and various authors have used 
different values for these. 
This probably accounts for the fact that the resummed calculations of 
Davies, Webber and Stirling\cite{dws}, 
and of Ellis and Veseli\cite{eandv} (also shown in the figure)
do not offer quite as good a description of the data.

D\O\ have also observed\cite{georg} the effect of QCD corrections
in the angular distribution of electrons from $W$ decay.
Figure~\ref{fig:georg} shows the extracted value of $\alpha_2$
(the coefficient of $\cos^2\theta^*$ in the angular 
distribution) as a function of $p_T^W$ .  The measurement is in
good agreement with QCD.

\subsection{$W+$jets}

QCD also predicts the number and spectrum of jets produced together
with the vector boson.
D\O\ used to show a cross section ratio $(W+1{\rm jet})/(W+0{\rm jet})$
which was badly in disagreement with QCD.  This is no longer shown:
the data were basically correct,  but there was a bug in the way
D\O\ extracted the ratio from the DYRAD theory calculation.

\begin{figure}[tb]
\begin{center}
\begin{tabular}{cc}
\includegraphics*[height=6.5cm]{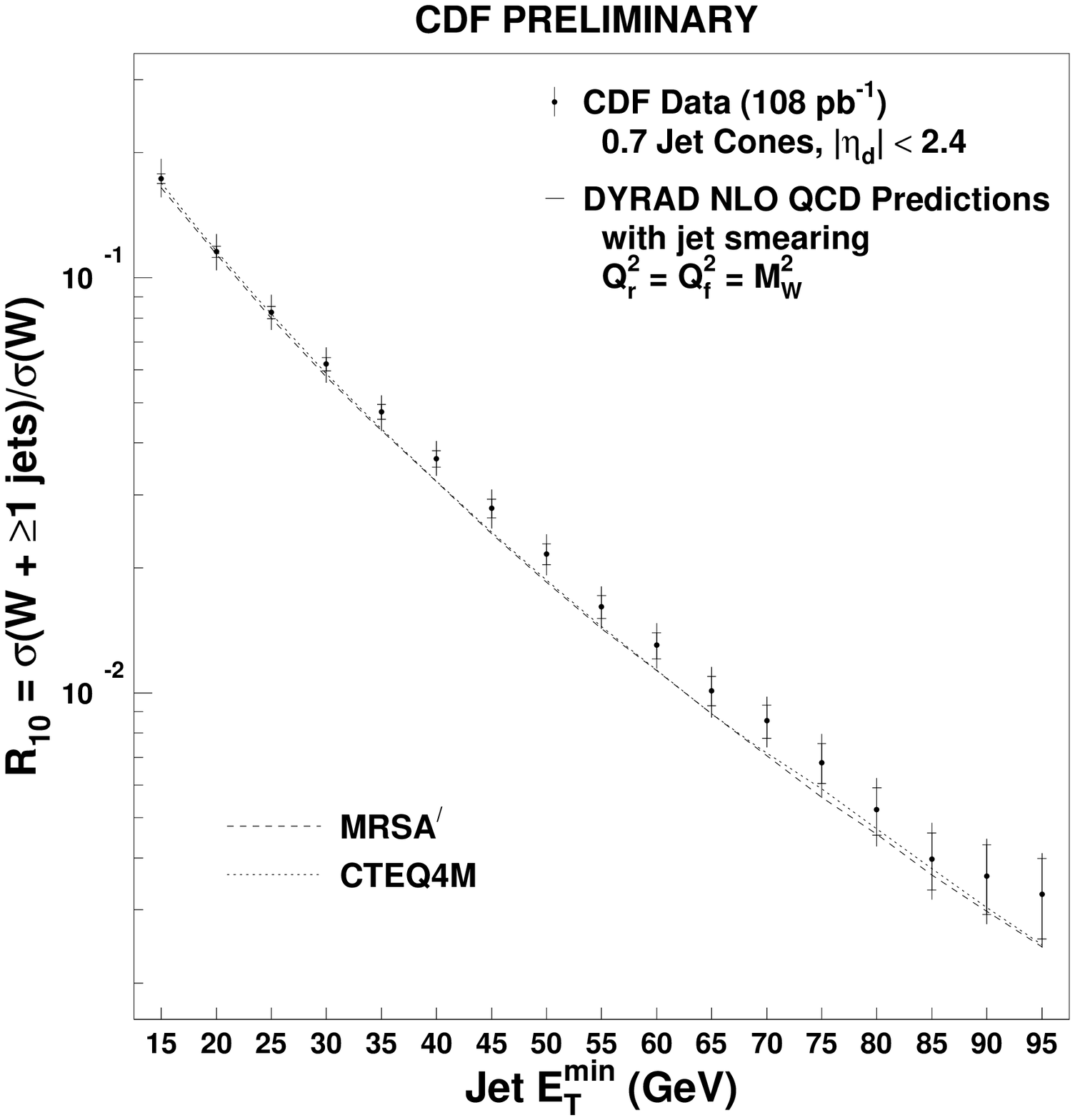}&
\includegraphics*[height=6.5cm]{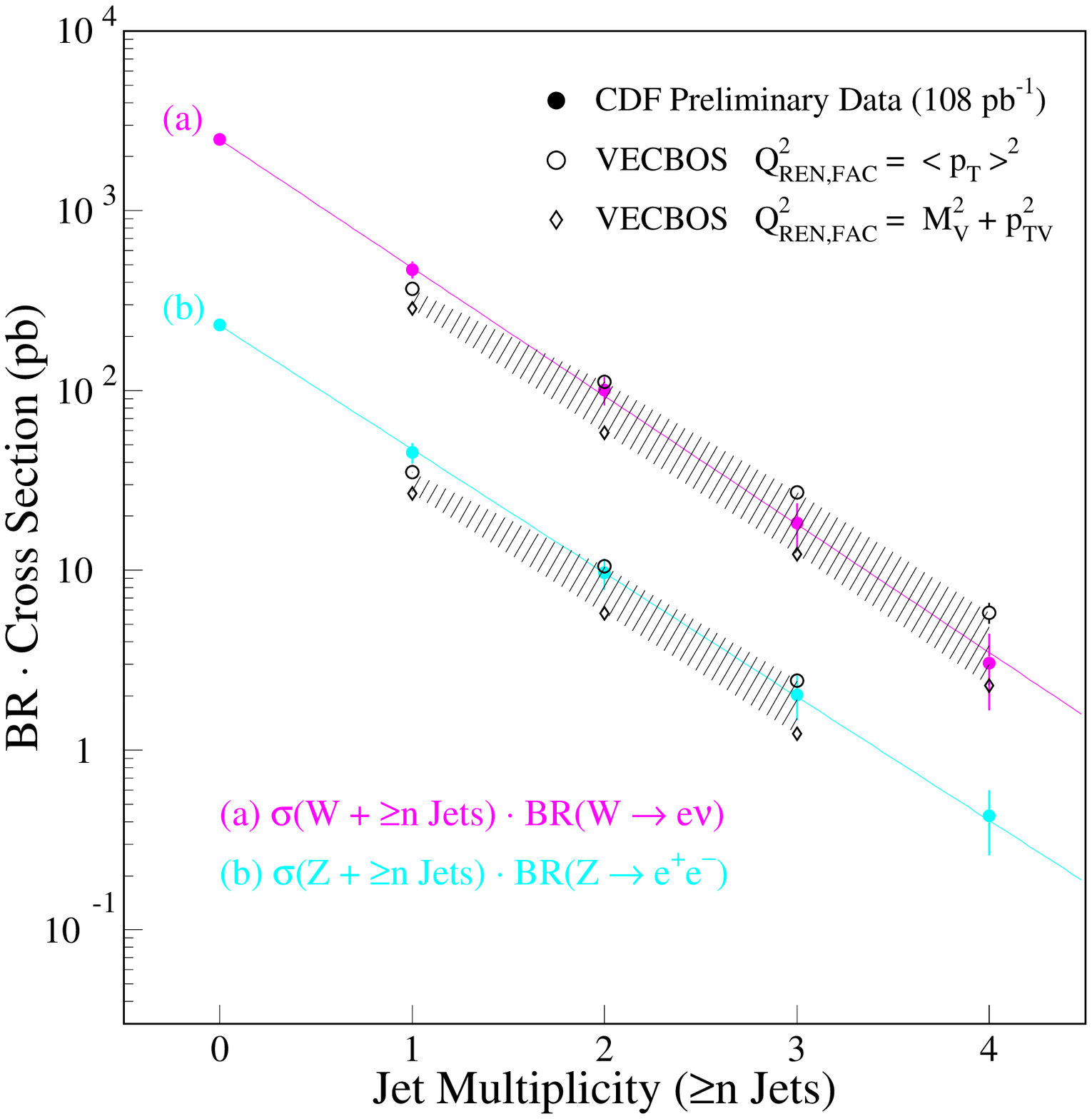}\\
\end{tabular}
\caption{The left hand plot shows the ratio of 
the $W+$~(one or more) jets cross section to the inclusive
$W$ cross section at the Tevatron, as measured by 
CDF\protect\cite{cdfwjets}.  The right
hand plot shows the $W+n$jets  and $Z+n$jets cross sections 
as a function of the number of jets $n$.
\label{fig:cdfwjets}}
\end{center}
\end{figure}

Recent CDF measurements of the $W+$jets cross sections\cite{cdfwjets} 
agree well with
QCD, as shown in Fig.~\ref{fig:cdfwjets}.  The figure shows
the fraction of $W$'s with 1 or more jets, compared 
with the NLO prediction; and the $W + n$~jets rate,
compared with the LO prediction (for a variety of renormalization
scales).  Alas, it seems that there is little prospect for being
able to extract $\alpha_s$ from these measurements, as had been 
hoped. This is because the $W+$jet cross section depends 
on $\alpha_s$ both in the jet production
vertex and in the parton distributions, and these two factors
largely cancel in the kinematic range probed at the Tevatron.  

\begin{figure}[p]
\begin{center}
\begin{tabular}{cc}
\includegraphics*[height=6.5cm]{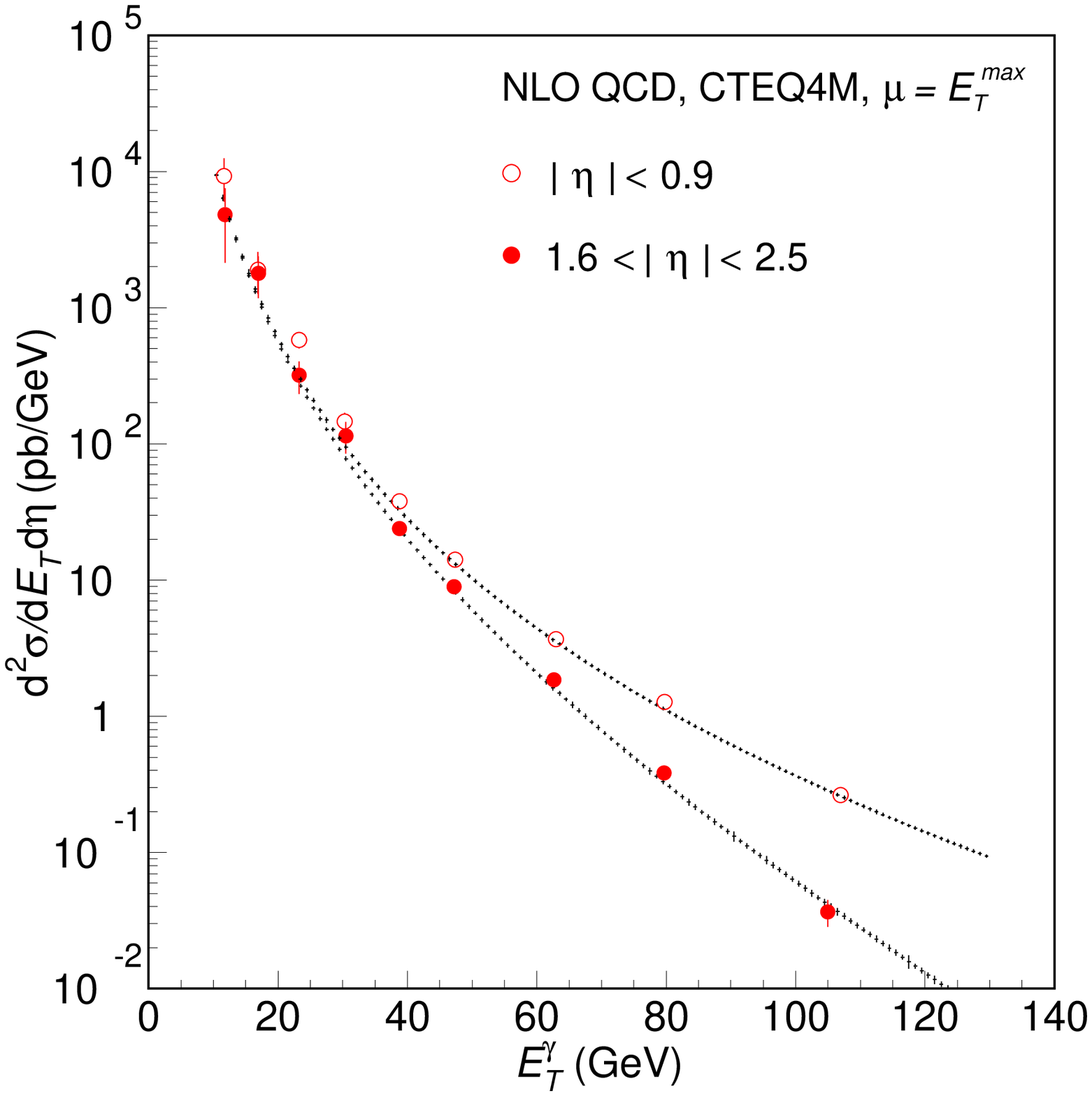}&
\includegraphics*[bb=30 140 550 655,height=6cm]{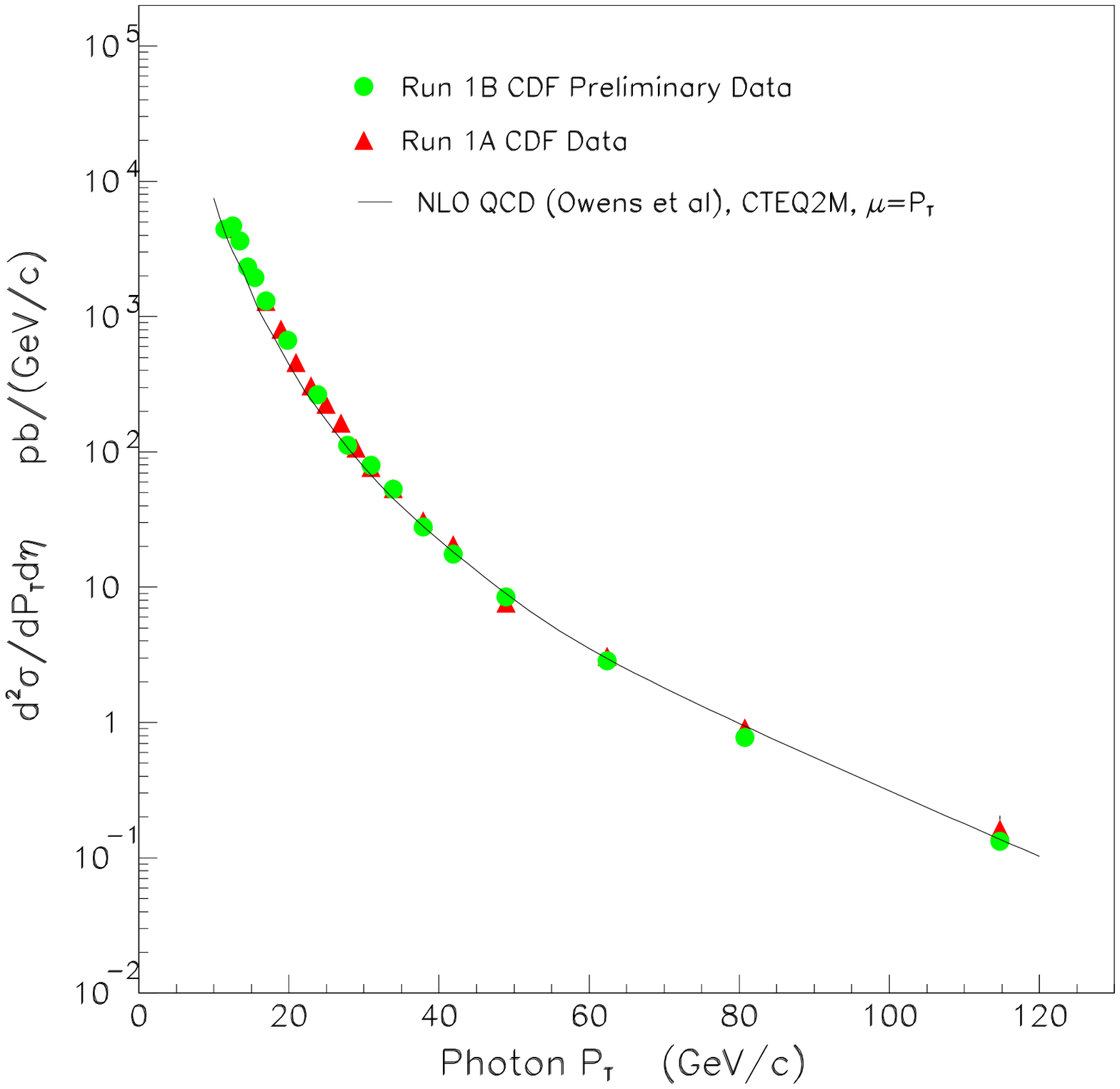}\\
\includegraphics*[height=6.5cm]{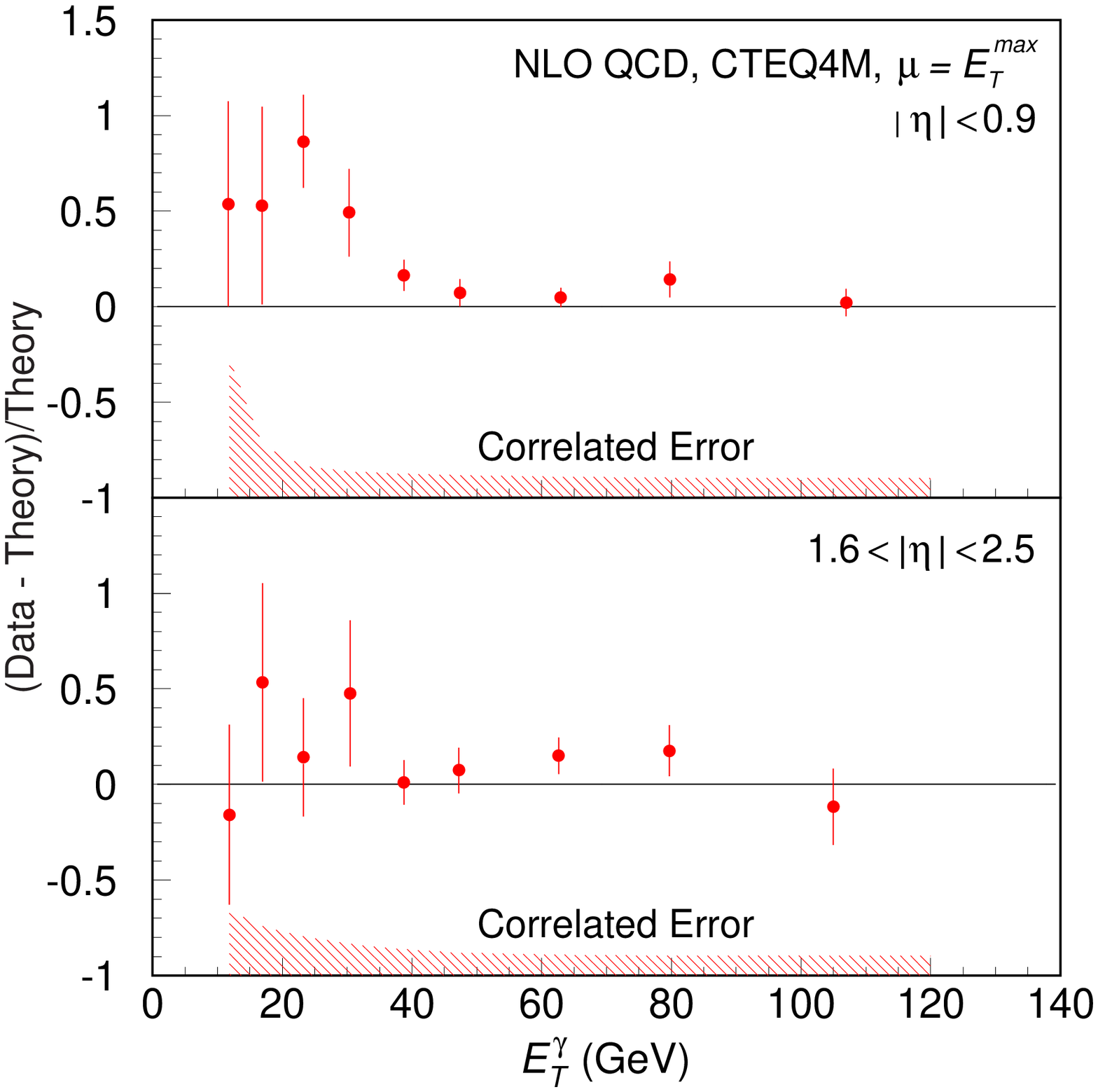}&
\includegraphics*[bb=30 140 550 655,height=6cm]{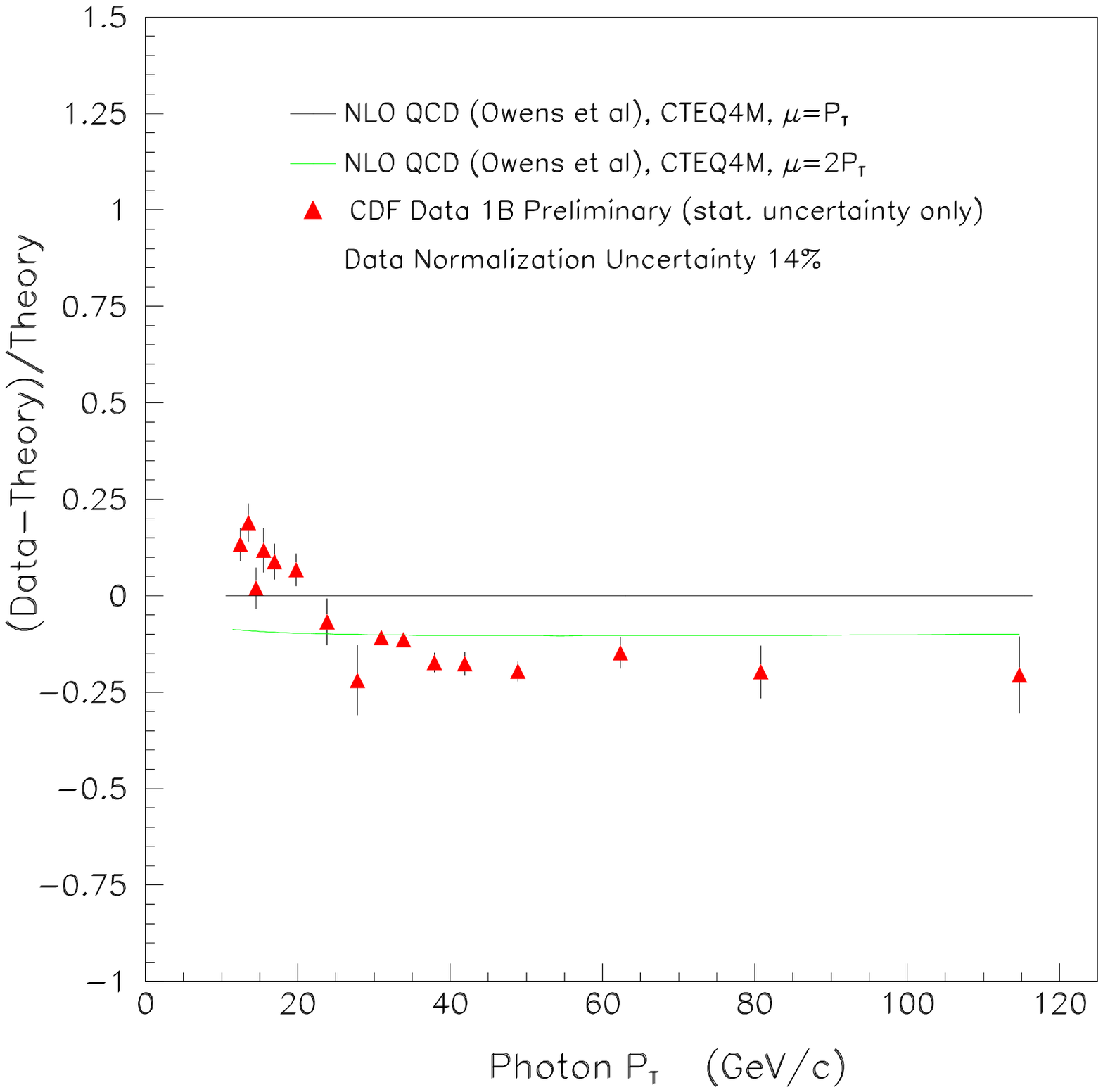}\\
\end{tabular}
\caption{Inclusive isolated direct photon cross sections at the Tevatron;
the left hand plots show D\O\protect\cite{d0photons} measurements 
and the right hand plots
show the latest CDF results\protect\cite{cdfnewphotons} 
(statistical errors only).  All are compared
with the NLO QCD prediction of Owens {\it et al}.\protect\cite{owens}.
\label{fig:photons}}
\end{center}
\end{figure}

\section{Isolated Photon Production}

Historically, many authors hoped that measurements of direct (or prompt) 
photons would provide a clean test of QCD, free from the 
systematic errors associated
with jets, and would help pin down parton distributions.  In fact photons
have not lived up to this promise --- instead they revealed that 
there may be unaccounted-for effects in QCD cross sections at 
low $E_T$.  (Because photons can typically be measured 
at lower energies than jets, they provide a way of exploring the
low-$E_T$ regime).  Results from the Tevatron experiments
\cite{d0photons}\cite{cdfnewphotons} are shown in 
Fig.~\ref{fig:photons}. While the general agreement with the NLO calculation
of Owens and collaborators\cite{owens} is good,
there is a definite tendency for the data to rise above the theory
at low transverse energies.  

\begin{figure}[tb]
\begin{center}
\begin{tabular}{cc}
\includegraphics*[bb=60 140 525 655,height=6cm]{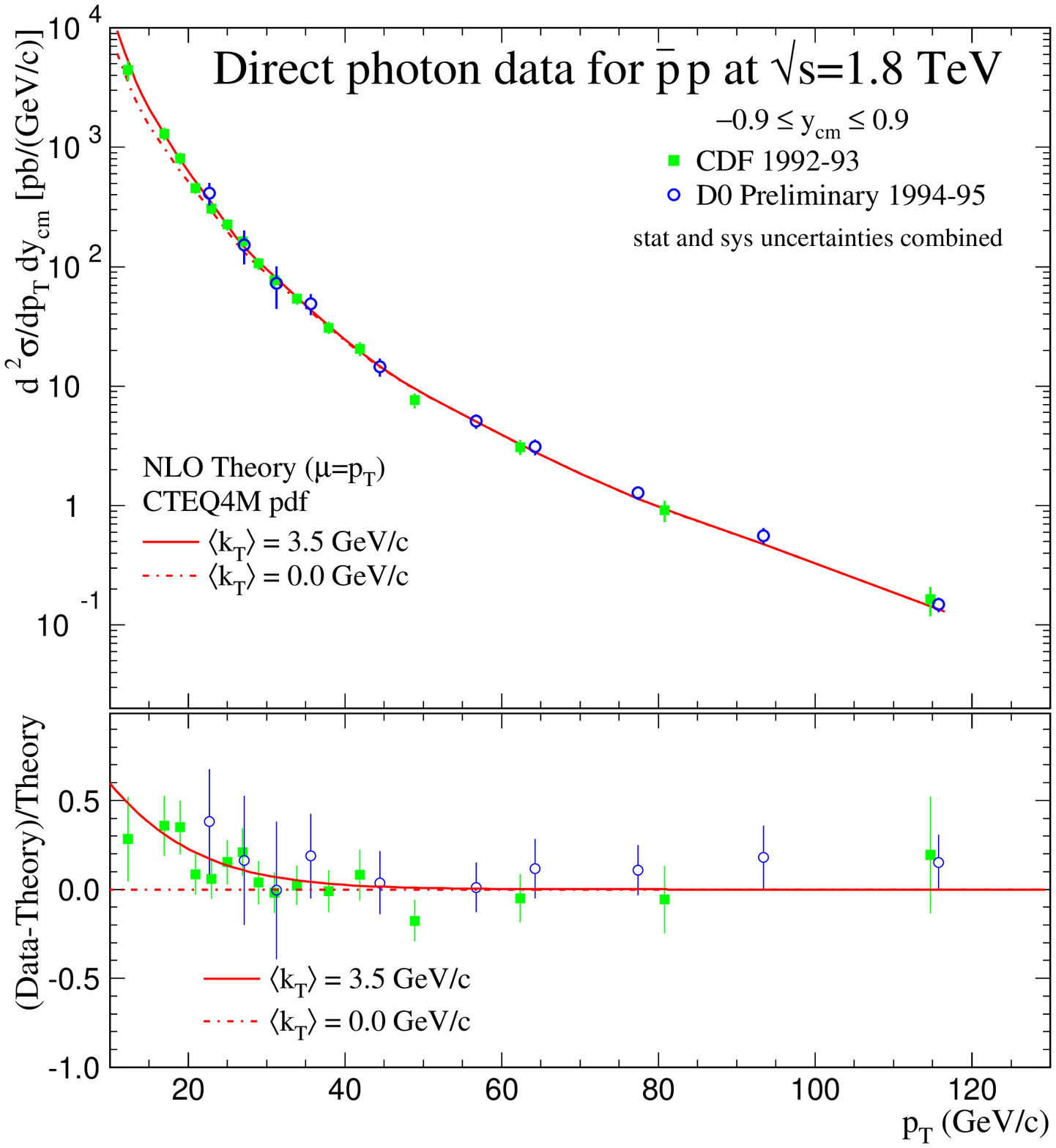}&
\includegraphics*[bb=30 162 525 635,height=6cm]{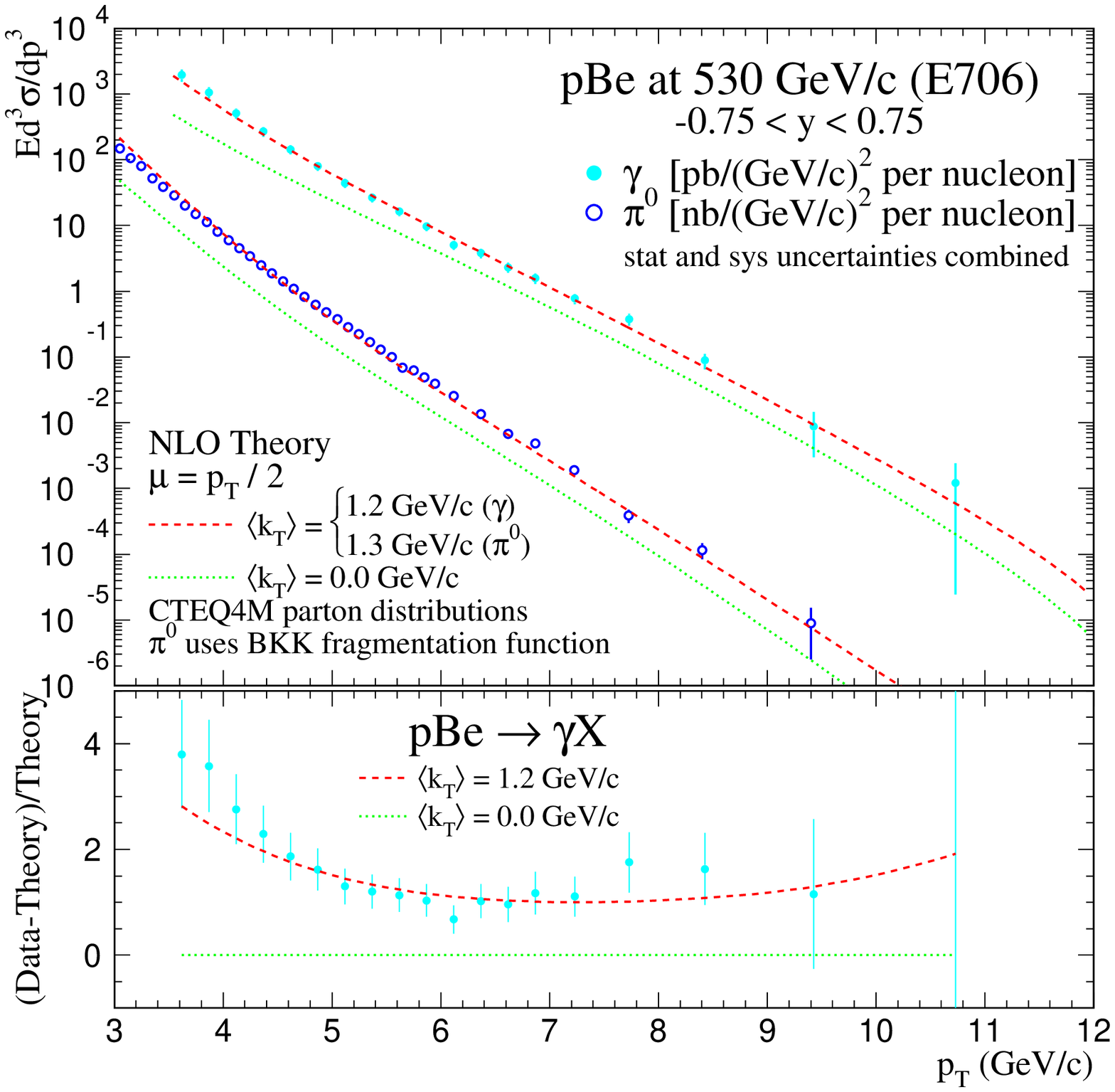}
\end{tabular}
\caption{The isolated photon cross section at the Tevatron (left hand plots)
showing the improved agreement
with QCD if 3.5~GeV of transverse momentum smearing 
(``$k_T$'') is added to account for soft gluon emission.
The right hand plot shows the isolated photon and $\pi^0$ cross 
sections measured by E706\protect\cite{e706}, 
compared with the NLO QCD prediction with and 
without 1.2~GeV of additional $k_T$ smearing. 
\label{fig:kttev}}
\end{center}
\end{figure}
\begin{figure}[p]
\begin{center}
\begin{tabular}{cc}
\includegraphics*[height=6cm]{e706_resum_comp.eps}&
\includegraphics*[height=6cm,angle=90]{laenenresum.eps}\\
\end{tabular}
\caption{Resummed calculations of isolated photon production compared
with the E706 data; 
left, by Catani {\it et al.}\protect\cite{cataniresum706} and 
right, by Laenen, Sterman and Vogelsang\protect\cite{laenenresum706}.
\label{fig:resum}}
\begin{tabular}{cc}
\includegraphics*[bb=20 50 600 700,height=7cm,angle=270]{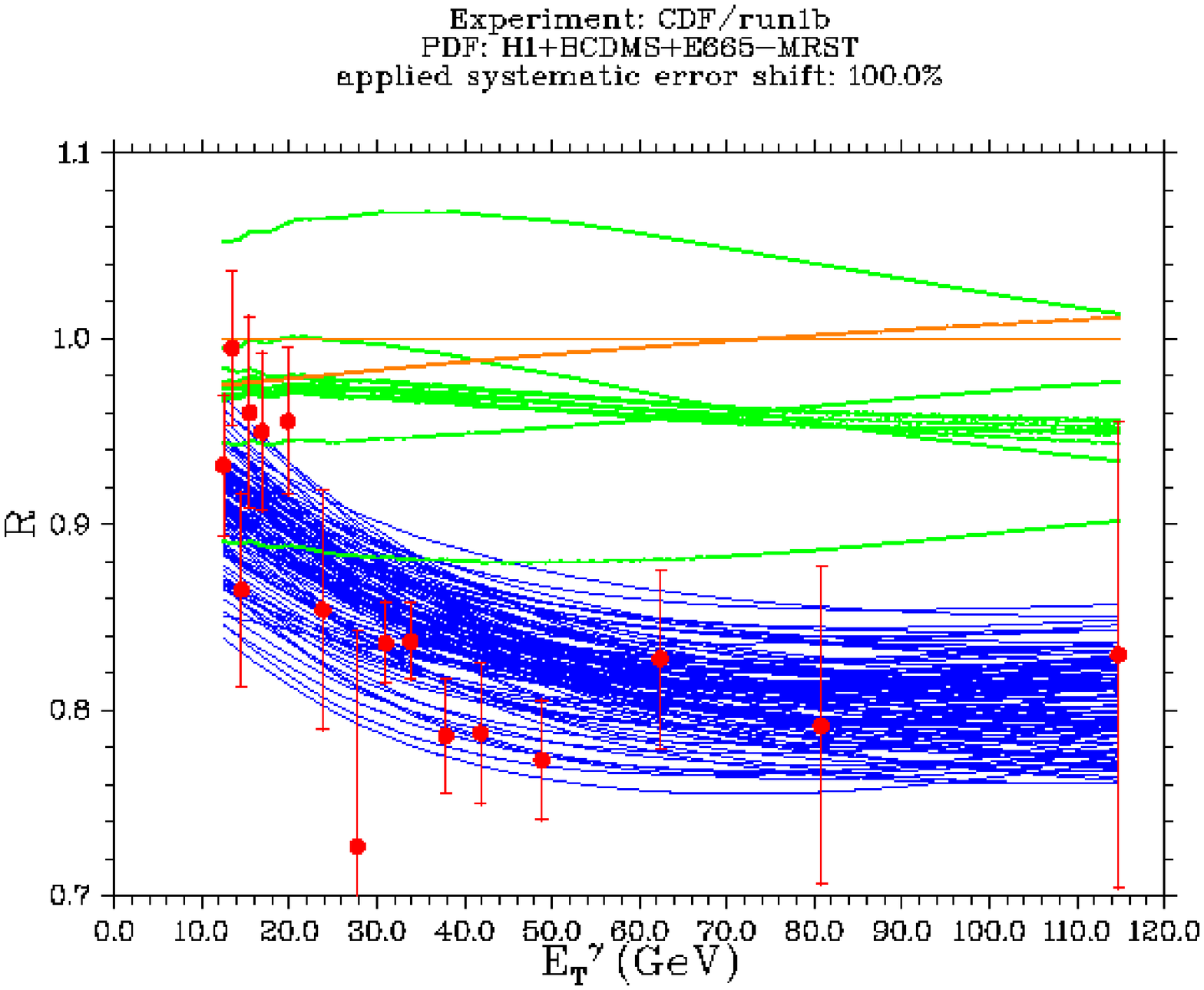}&
\includegraphics*[bb=20 50 600 700,height=7cm,angle=270]{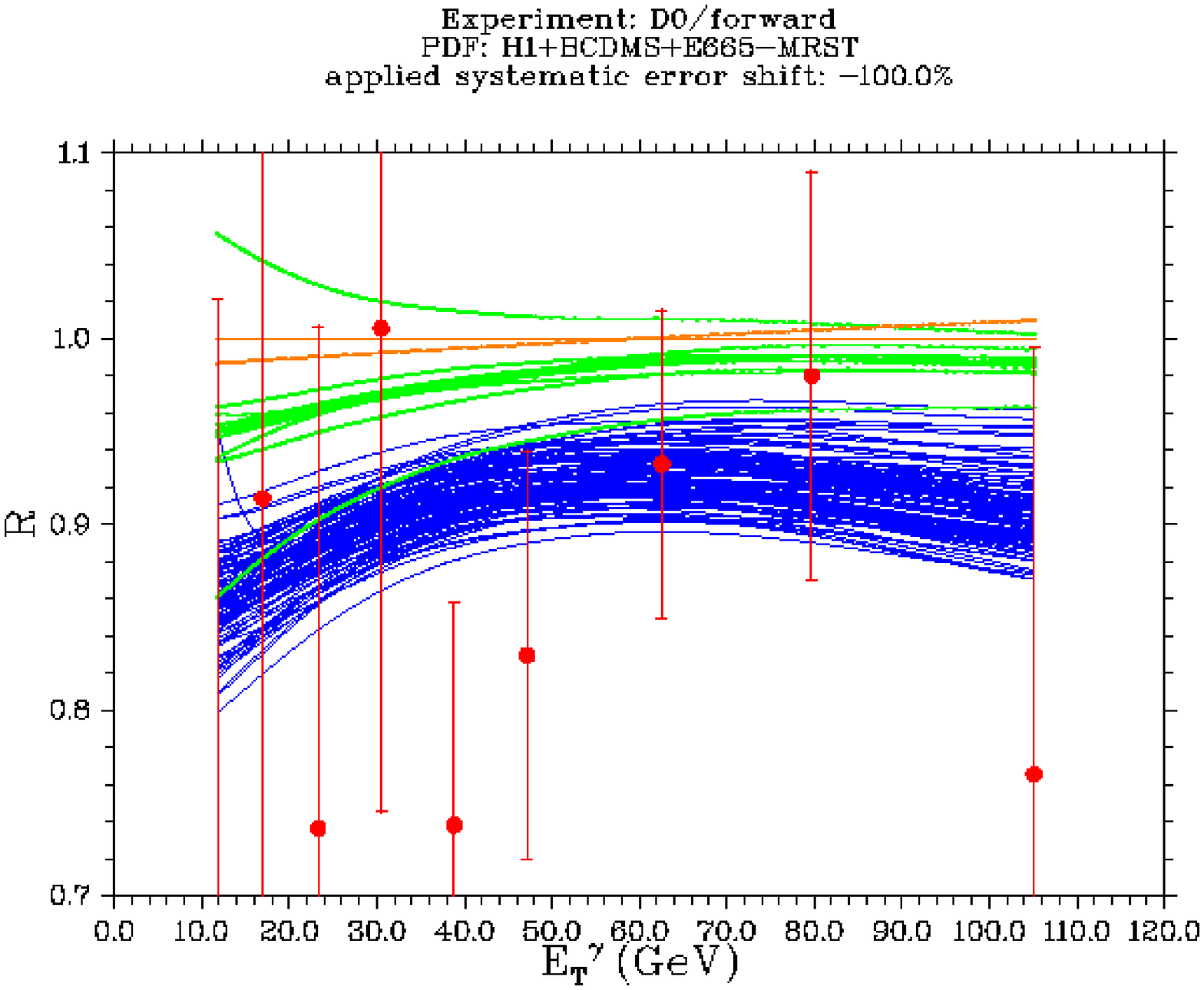}
\end{tabular}
\caption{Isolated photon cross sections 
measured in the central
region by CDF (left) and the forward region by D\O\ (right), 
compared with NLO
QCD predictions.  The blue curves use an ensemble of PDF's proposed by Giele,
Keller and Kosower, derived by fitting to H1, BCDMS and E665 data.
The range of predictions gives a measure of the uncertainty
on the PDF.  The green curves use MRS99 distributions and the orange
curves are CTEQ5M and 5L.  No additional $k_T$ smearing is included.
\label{fig:giele}}
\end{center}
\end{figure}


An often-invoked explanation for this effect is that there exists
additional
transverse momentum smearing of the partonic system due to soft gluon
radiation.  The magnitude of the smearing, or ``$k_T$'', is typically
a few GeV (at the Tevatron), motivated in part by the experimentally
measured $p_T$ of the $\gamma\gamma$ system in diphoton production 
which peaks around 3~GeV\cite{diphotons}.  PYTHIA simulations of
photon production  also suggest that the most probable transverse
momentum
of radiated initial state gluons is 2--3~GeV\cite{cdfnewphotons}.
Inclusion of such $k_T$ through Gaussian
smearing in the calculation gives much better agreement with the data,
as shown in Fig.~\ref{fig:kttev}.
Much larger deviations from QCD are observed in fixed-target experiments
such as E706 at Fermilab\cite{e706}.  Again, Gaussian smearing (with 
$k_T \approx 1.2$~GeV in this case) can account for the data, as also
shown in the figure.

Unfortunately the predictive power of
Gaussian smearing is small: it cannot really tell us what 
happens to forward photons, or what happens at the LHC,
for example.  The ``right way'' to treat soft gluon emission should be 
through a resummation calculation which works nicely for $\gamma\gamma$ and
$W/Z$ transverse momentum distributions.  Initial attempts
did not seem to model the E706 data\cite{cataniresum706}\cite{kidoresum706},
but more recent calculations include aditional terms and
look more promising\cite{laenenresum706}
(Fig.~\ref{fig:resum}).  

A rather different view is expressed by Aurenche and 
collaborators\cite{aurenche},
who find their calculations, {\it sans} $k_T$, to be consistent with all 
the ISR and fixed-target data 
with the sole exception of E706.  
They say, ``it does not appear very 
instructive to hide this problem by introducing an arbitrary parameter
fitted to the data at each energy,'' by which they mean $k_T$.  

The latest result in this saga is most interesting.  Elsewhere 
in these proceedings, Walter Giele reports that he is able to 
obtain good agreement (Fig.~\ref{fig:giele}) 
between the Tevatron data and QCD, without
any $k_T$, with a newly derived set of PDF's that are
extracted from DIS data from H1, BCDMS and E665.  
If correct, this observation could render the whole discussion
moot --- there would be no discrepancy with QCD here at all,
merely another indication that we need to understand parton
distribution uncertainties!  

\begin{figure}[tb]
\begin{center}
\begin{tabular}{cc}
\includegraphics*[height=6.5cm]{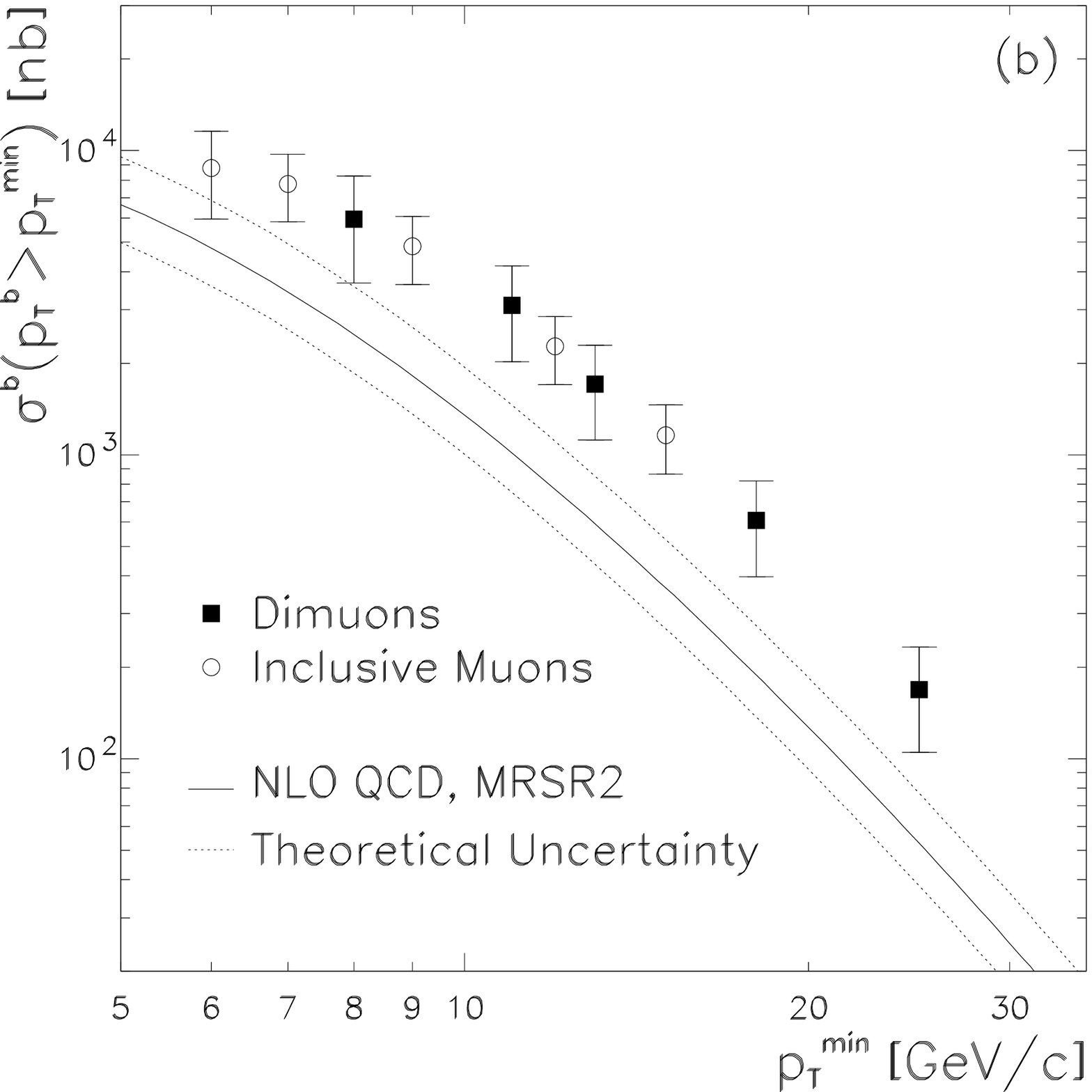}&
\includegraphics*[height=6.5cm]{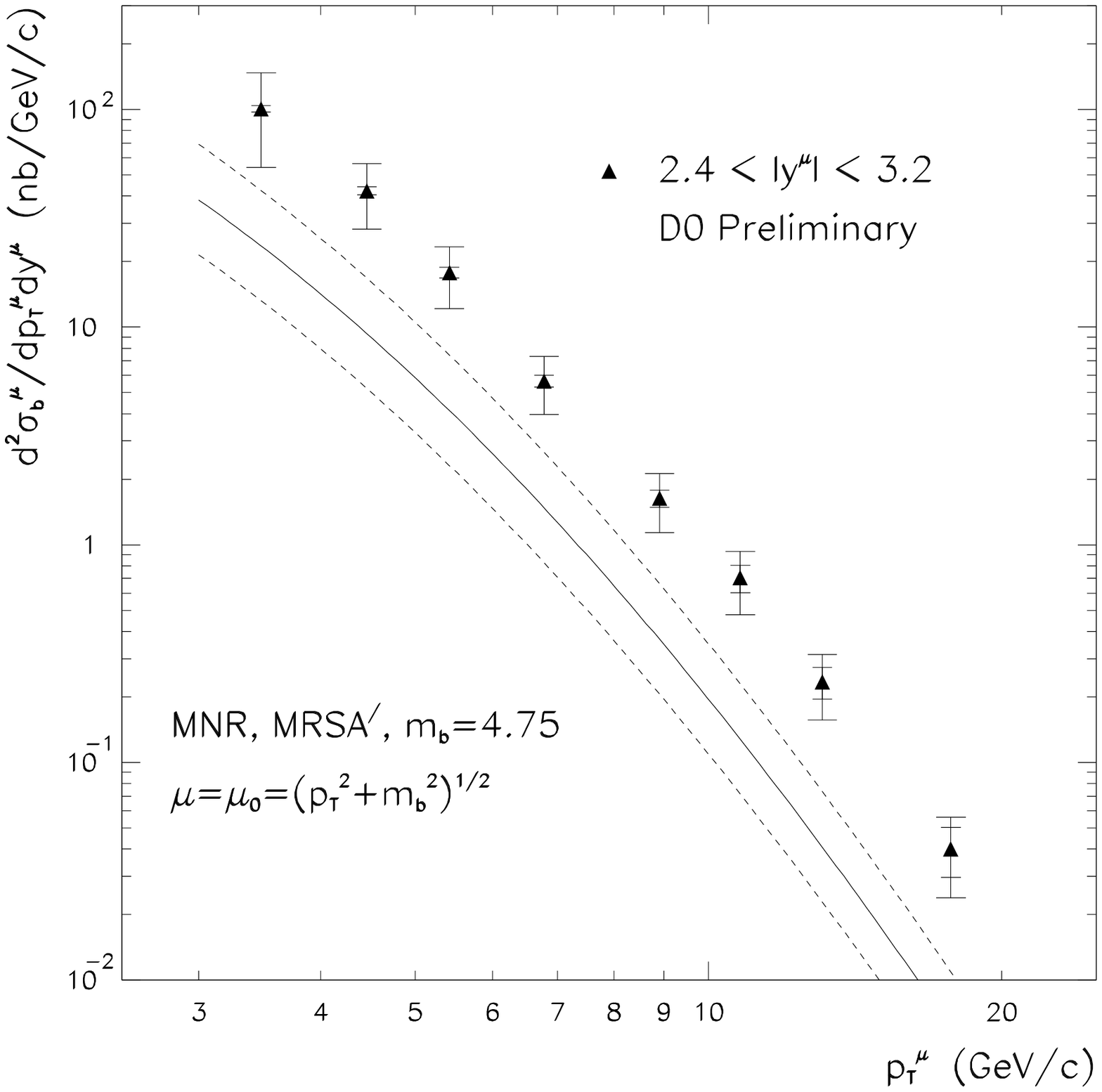}\\
\end{tabular}
\caption{Cross sections for $b$ production at the Tevatron compared
with NLO QCD predictions, as measured by D\O\protect\cite{d0b};
left, central rapidity region, and
right, forward.
\label{fig:d0b}}
\end{center}
\end{figure}

In summary, direct photon production has proved extremely interesting
and remains quite controversial. The appropriateness of a Gaussian $k_T$
treatment is still hotly debated, the experiments may not all be 
consistent, and the latest results merely increase the mystery --- is
it all just the PDF's?

\section{Heavy Flavour Production}

\begin{figure}[tb]
\begin{center}
\begin{tabular}{cc}
\includegraphics*[height=6.5cm]{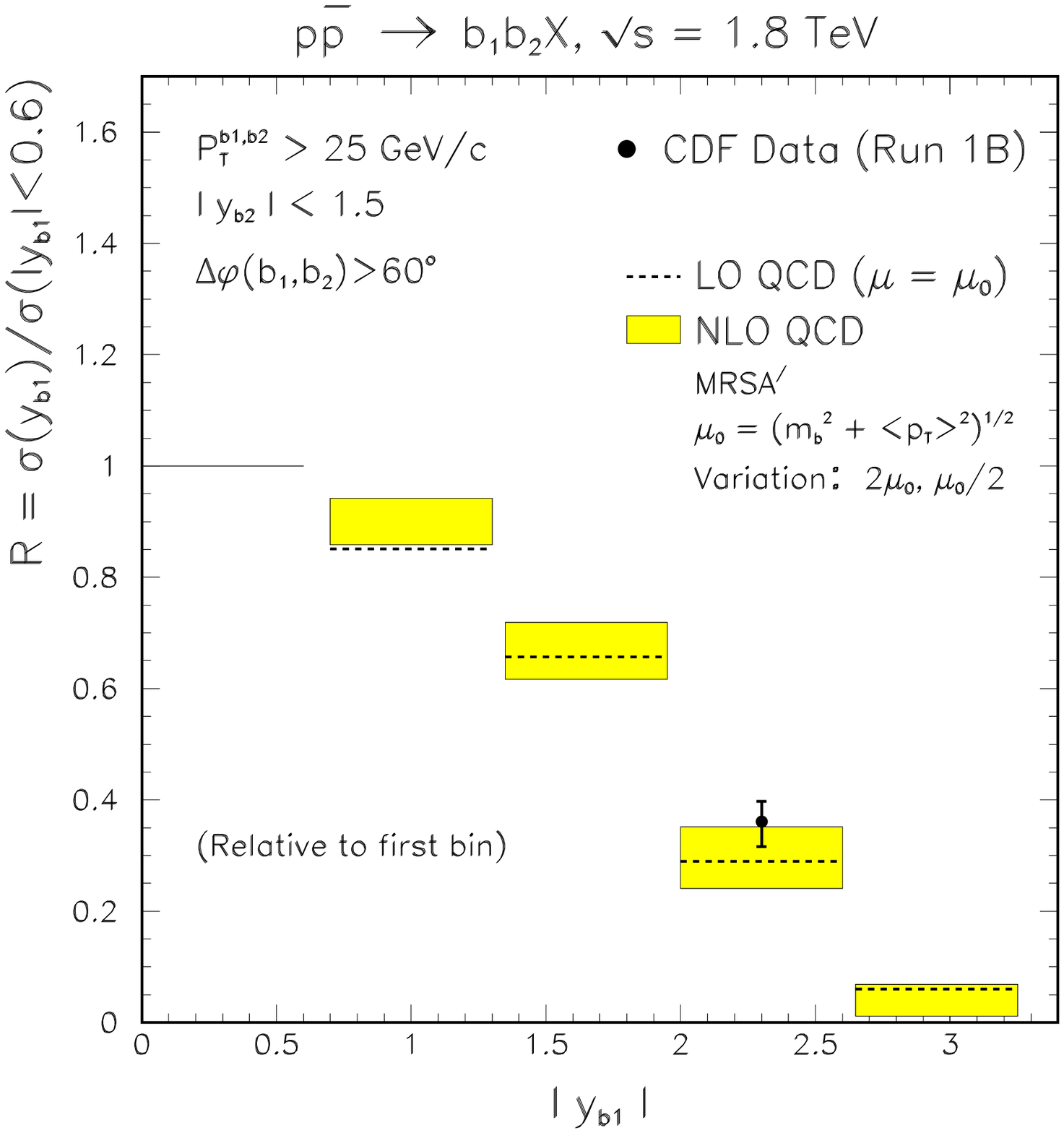}&
\includegraphics*[height=6.5cm]{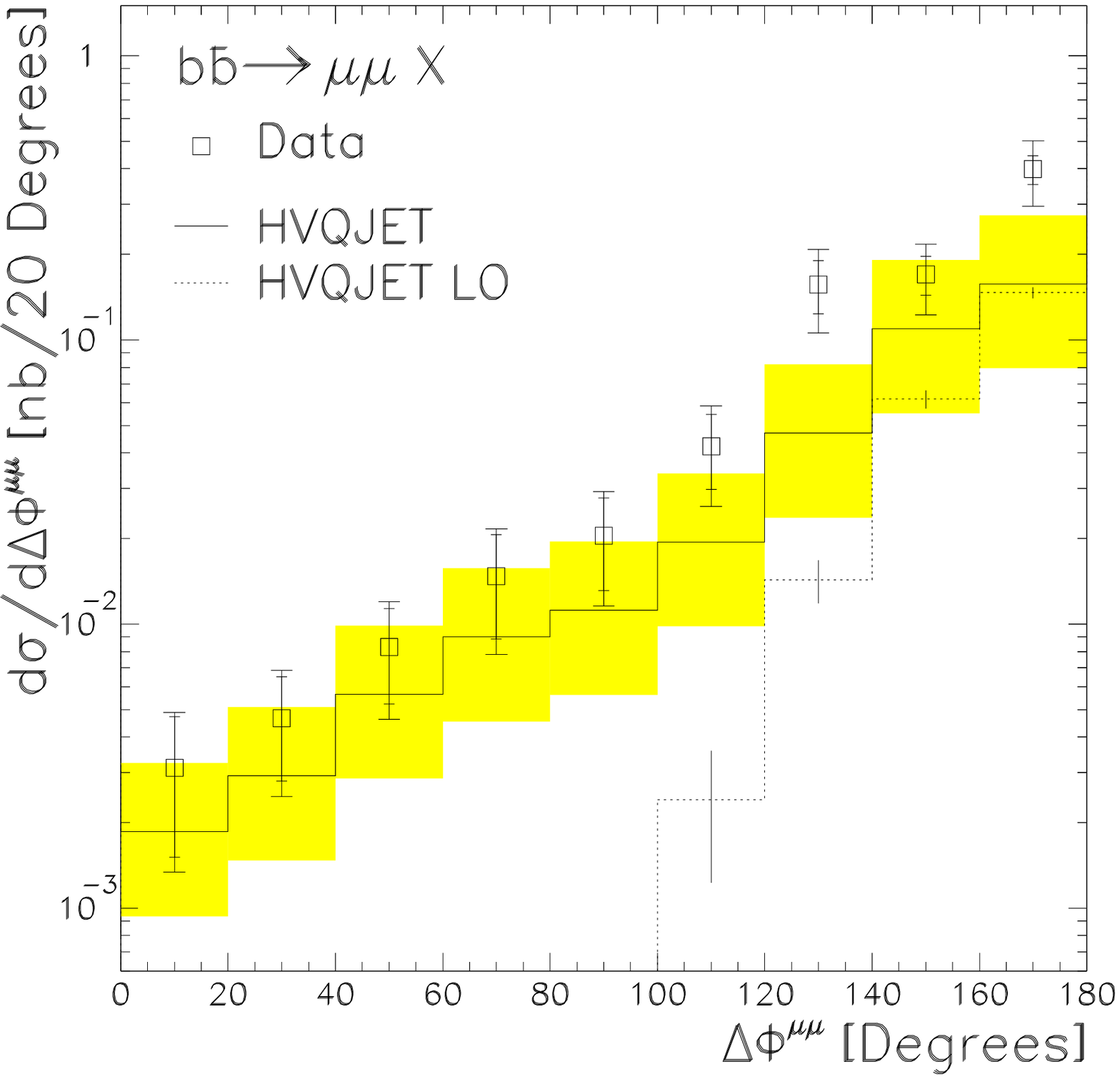}\\
\end{tabular}
\caption{Correlations between $b$-jets at the Tevatron compared
with NLO QCD predictions; (left) 
rapdidity correlations as measured by CDF (normalized to the
first bin), and (right)
azimuthal angle correlations as measured by D\O.
\label{fig:bcorrel}}
\end{center}
\end{figure}

\begin{figure}[tb]
\begin{center}
\begin{tabular}{cc}
\includegraphics*[bb=30 140 525 655,height=6cm]{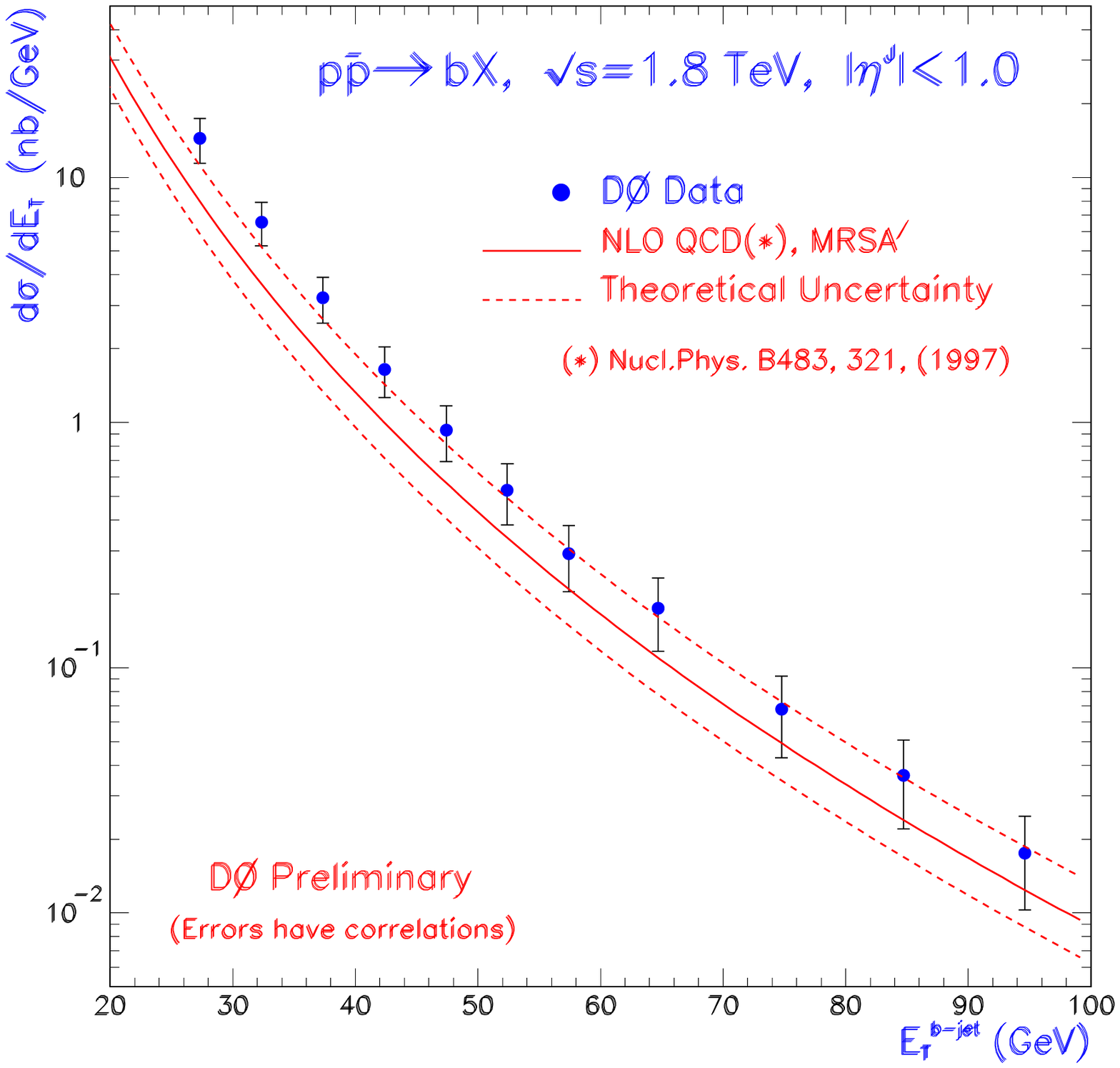}&
\includegraphics*[bb=30 140 525 655,height=6cm]{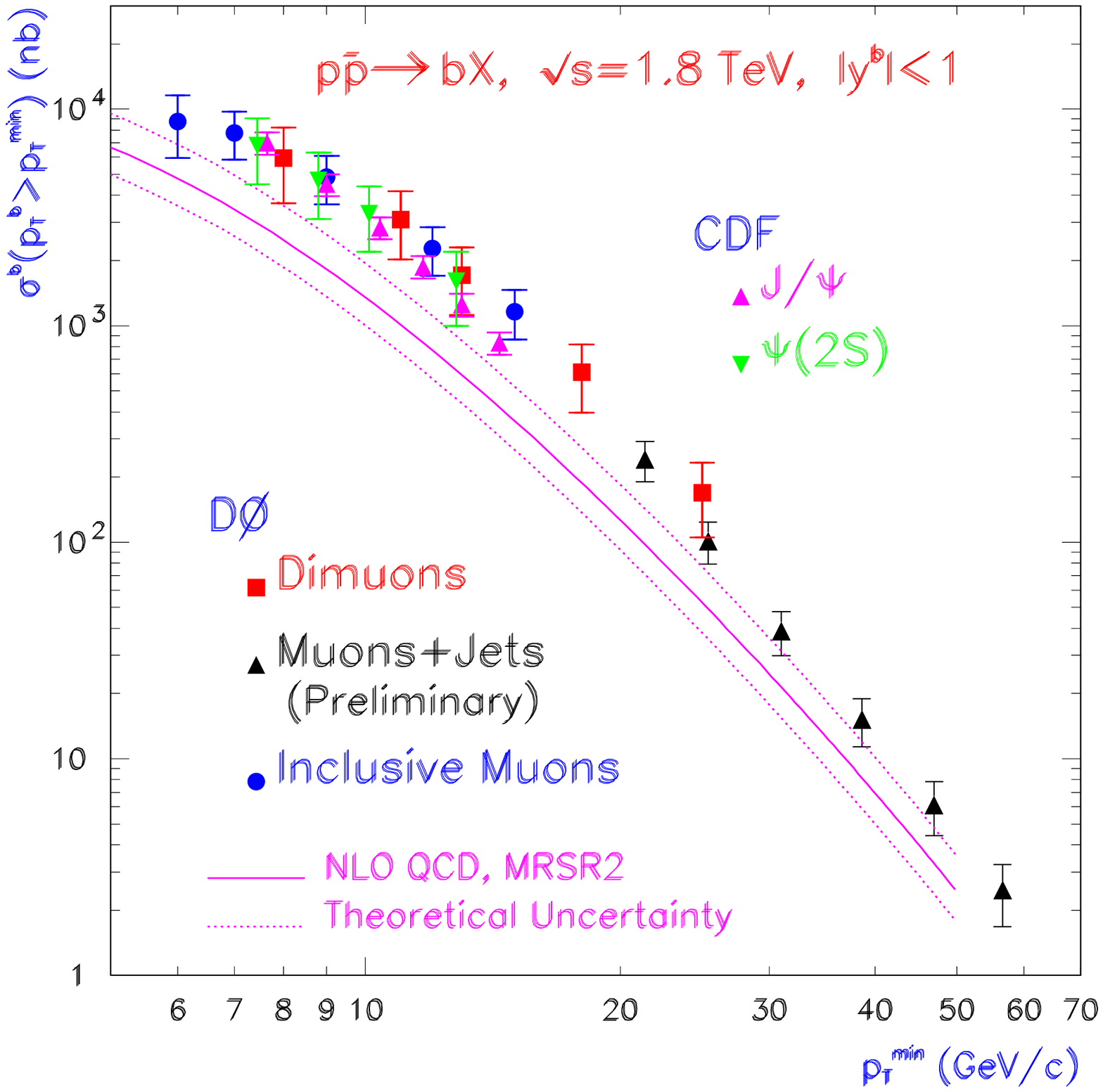}\\
\end{tabular}
\caption{New measurements of high-$p_T$ $b$-jet production at 
the Tevatron compared with NLO QCD predictions, as measured 
by D\O\protect\cite{d0bnew}; (left) differential cross
section as a function of $E_T^{\rm jet}$; 
(right) integral cross section as a function of $p_T^b$.
\label{fig:d0bnew}}
\end{center}
\end{figure}

At the Tevatron, the measured inclusive $b$-quark 
and $B$-meson production cross
sections continue to lie a factor of about two above the NLO QCD 
expectation.  This is seen by both CDF\cite{cdfb} and D\O\cite{d0b} 
in the central and
forward regions (the difference is perhaps even larger for
forward $b$ production, as seen in Fig.~\ref{fig:d0b}).  
On the other hand, NLO QCD does a good job
of predicting the shape of inclusive distributions, and of the
correlations between $b$ quark pairs (Fig.~\ref{fig:bcorrel}), 
so it seems unlikely that
any exotic new production mechanism is responsible for the higher
than expected cross section.  
In passing,
it is interesting to note that a similar excess is also seen in
$b$-production at HERA\cite{h1bquark}\cite{zeusbquark} 
and in $\gamma\gamma$ collisions at LEP2\cite{l3bquark}.

\begin{figure}[tb]
\begin{center}
\includegraphics*[width=\textwidth]{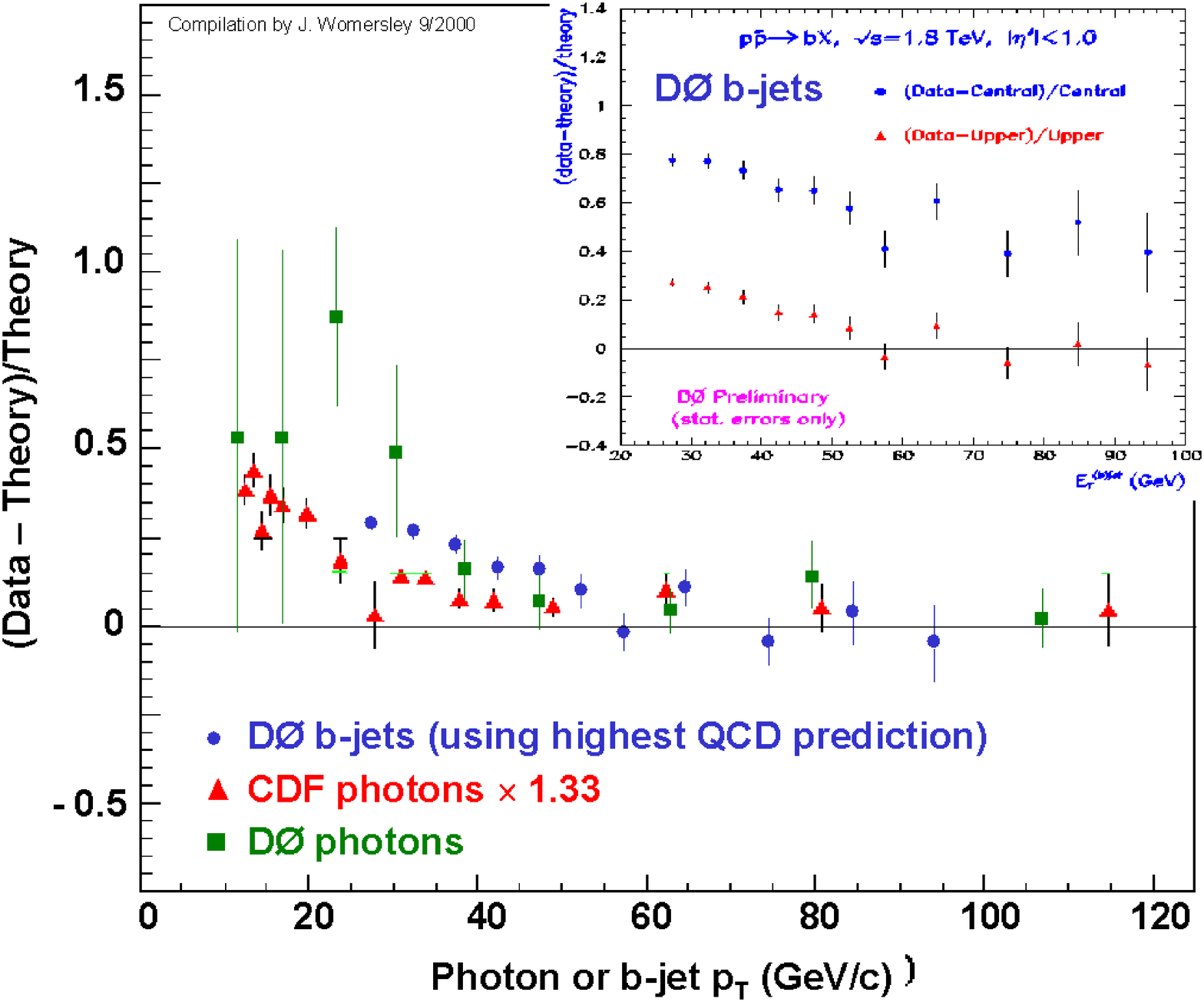}
\caption{Compilation of $b$-jet and isolated photon cross sections
compared with QCD.
\label{fig:bjets}}
\end{center}
\end{figure}

Recently, D\O\ have extended these measurements to higher transverse
momenta (up to 100~GeV)\cite{d0bnew}.
The results (Fig.\ref{fig:d0bnew}) are interesting:  the measured
cross section comes closer to the prediction around $p_T \sim 50$~GeV
and above.  It is therefore tempting to compare the shape of 
{\it(Data$-$Theory)/Theory} for $b$-jets and for
photons, as I have done in Fig.~\ref{fig:bjets}.  
The plot compares D\O\ photons, CDF photons (renormalized
by 1.33) and D\O\ $b$-jets (compared with the highest of the range
of QCD predictions).
The plot is perhaps quite suggestive that the same explanation
may be relevant for photons and $b$-jets --- whatever that may
be!

If the heavy flavour is heavy enough, QCD seems to work rather better.
The current state of measured and predicted top cross sections is
summarised in Table~\ref{tab:top}.  This includes the latest (revised)
CDF measurement.  There is an excellent agreement between data and
theory, though one may note that the most recent (resummation) 
calculation\cite{kidonakistop} lies outside the band of uncertainty
claimed by earlier authors\cite{bergertop}.  

\index{top}

\begin{table}[tb]
\begin{center}
\begin{tabular}{|l|c|}  
\hline
Authors & Cross Section (pb)\\
\hline
\hline
CDF\protect\cite{cdftop}  & $6.5^{+1.7}_{-1.4}$ (at $m_t=175$~GeV)\\
D\O\protect\cite{d0top}   & $5.9\pm 1.7$ (at $m_t=172$~GeV)\\
\hline  
Bonciani {\it et al.}\protect\cite{boncianitop} & $5.0\pm 1.6$\\
Berger and Contopanagos\protect\cite{bergertop} & $5.6^{+0.1}_{-0.4}$\\
Kidonakis\protect\cite{kidonakistop} & $6.3$\\
\hline
\end{tabular}
\caption{Top production cross sections at the Tevatron, measured and
predicted.}
\label{tab:top}
\end{center}
\end{table}

\section{Prospects}

What can we look forward to in Run~2 and beyond?  
Clearly there will be lots more data --- the next decade belongs to the hadron
colliders.  We can also expect improved calculations (NNLO calculations,
NLL resummations).  There is a lot of work going on towards the
goal of correctly treating uncertainties in PDF's, as manifested
by Walter Giele's contribution to these proceedings.  This is a great
step forward, but it does impose significant work on the experiments,
who must understand and publish all the errors and their correlations.

We can also look forward to improved jet algorithms.  There is a CDF-D\O\
accord from the Fermilab Run~2 QCD workshop.  
The $k_T$ algorithm will be used
from the start, and the experiments have agreed upon one common 
implementation.  They will also try to make the cone algorithm theoretically
more acceptable by modifying the choice of seeds (or even through
a seedless version). 

I would also like to see a theoretical and experimental effort to
understand the underlying event, and include it in the predictions.
The current approach is to subtract an ``underlying event contribution''
from the jet energies.  This assumes factorization between the
hard scattering and the underlying event, and while this is a
reasonable approximation it is bound to break down at the GeV
level because the hard event and the underlying event are 
color-connected.
Indeed, HERWIG suggests that at the 1--2 GeV level 
jets pick up or lose energy to the rest of the event depending 
on the jet algorithm.  This is another example of how greater
precision demands greater care: approximations and
assumptions that used to be ``good enough'' can no longer
be taken for granted. 
In fact there are very nice new results 
from CDF on the underlying event\cite{cdfule}.  
Understanding the underlying event would
also allow a consistent treatment of double parton scattering.

Finally, one may hope for a consistent approach to hard 
diffraction processes.  High $E_T$ jets and $W$ production
are hard processes which should be amenable to perturbative
calculation, even if the final state is such that one of the
nucleons does not break up.  We need to break down the walls 
of the ``pomeron ghetto'' and stop trying to describe these
processes in a language which, in my opinion, does not 
promote understanding.

\section{Conclusions}

Tevatron QCD measurements have become precision measurements.
We are no longer testing QCD; we are testing our ability to
make precise calculations within the framework of QCD. The
state of the art is NNLO calculations, NLL resummations, and
measurement errors at the 5\% level.  This level of precision
demands considerable care both from the experimentalists
and the phenomenologists, in understanding jet algorithms,
jet calibrations, all the experimental errors and their
correlations, and the level of uncertainty in the calculations
and in the PDF's.  

In general our calculational tools are working well. The open
issues generally relate to attempts to push calculations closer
to the few-GeV scale ($b$ production at modest $p_T$, 
perhaps low-$E_T$ photons) and/or to regions where the
parton distributions are uncertain (high-$E_T$ jets, and
perhaps photons).

\Acknowledgments
I am grateful to Georgia Hamel and Jacqueline Pizzuti at SCIPP for
their efficient administration of the conference, and
to Howie Haber, Stan Brodksy and the other organizers of RADCOR-2000 
for making the program both informative and enjoyable.

\end{document}